\documentclass[preprint,onecolumn,nofootinbib]{revtex4}
\pdfoutput=1
\usepackage{CJK}
\usepackage[colorlinks=true,linkcolor=blue,urlcolor=blue,filecolor=black,citecolor=red,pdfstartview=FitV,pdftitle={},pdfsubject={},pdfkeywords={},pdfpagemode=None,bookmarksopen=true]{hyperref}
\usepackage{graphicx}
\usepackage{amsmath}
\usepackage{amsfonts}
\usepackage{amssymb,ulem}
\usepackage{color}%
\usepackage{dcolumn}
\newcommand{\f}{\begin{equation}}
\newcommand{\ff}{\end{equation}}
\newcommand{\fa}{\begin{eqnarray}}
\newcommand{\ffa}{\end{eqnarray}}

\begin{document}
\title{Higher derivatives driven symmetry breaking in holographic superconductors}
\author{Hai-Li Li$^{1}$}
\author{Guoyang Fu$^{2}$}
\author{Yan Liu$^{1}$}
\author{Jian-Pin Wu$^{2}$}
\thanks{jianpinwu@yzu.edu.cn}
\author{Xin Zhang$^{1,3,4}$}
\thanks{zhangxin@mail.neu.edu.cn}
\affiliation{
$^1$ Department of Physics, College of Sciences, Northeastern University, Shenyang 110819, China\ \\
$^2$ Center for Gravitation and Cosmology, College of Physical Science and Technology, Yangzhou University, Yangzhou 225009, China\ \\
$^3$ Ministry of Education's Key Laboratory of Data Analytics and Optimization for Smart Industry, Northeastern University, Shenyang 110819, China\ \\
$^4$ Center for High Energy Physics, Peking University, Beijing 100080, China}
\begin{abstract}

In this paper, we construct a novel holographic superconductor from higher derivative (HD) gravity
involving a coupling between the complex scalar field and the Weyl tensor.
This HD coupling term provides a near horizon effective mass squared, which can violates IR Breitenlohner-Freedman (BF) bound
by tuning the HD coupling and induces the instability of black brane such that the superconducting phase transition happens.
We also study the properties of the condensation and the conductivity in the probe limit.
We find that a wider extension of the superconducting energy gap ranging from $4.6$ to $10.5$
may provide a novel platform to model and interpret the phenomena in the real materials of
high temperature superconductor.

\end{abstract}
\maketitle
\section{Introduction}

Based on AdS/CFT (Anti-de Sitter/Conformal Field theory) correspondence \cite{Maldacena:1997re,Gubser:1998bc,Witten:1998qj,Aharony:1999ti},
a holographic superconductor model is suggested in \cite{Hartnoll:2008vx}.
In this model, a complex charged scalar field is introduced in Schwarzschild-AdS (SS-AdS) black brane to spontaneously break the U(1) symmetry
and transform to a charged scalar hair black brane.
The charged scalar field in the bulk is dual to the ``Cooper pair" operator at the boundary and the vacuum expectation value is the order parameter.
The symmetry breaking is introduced by a negative mass squared of the scalar field, which is allowed due to the Breitenlohner-Freedman (BF) bound in AdS spacetime \cite{Hartnoll:2008vx,Gubser:2008px}.
In \cite{Kim:2009kb}, a positive mass squared situation was studied. It was shown that as $m^{2}$ increase the phase space folds due to the non-linearity of the equations of motion,
so the two nearby points in the phase space can represent symmetry breaking. And they show that for a small positive mass squared,
the results is not much different from the negative case which has been studied in \cite{Horowitz:2008bn,Basu:2008st,Herzog:2008he,Hartnoll:2008kx,Gubser:2008wv}.

In \cite{Denef:2009tp}, they discuss the superconductivity instability by studying the normalisable solution of equations of motion (EOMs) for the charged scalar field on top of RN-AdS black brane geometry.
In essence, it is that the near horizon effective mass squared is below the AdS$_2$ BF bound, which induces the instability.
At the same time, we also require that the mass squared is above the boundary AdS$_4$ BF bound, which guarantees the stability of scalar field at the boundary.
In this paper, we construct a novel holographic superconductor model by introducing a higher derivative (HD) term, which couples the scalar field and Weyl tensor.
This HD term provides a near horizon effective mass squared but doesn't modify the boundary AdS$_4$ BF bound.

The superconducting energy gap is an important characteristic of superconductor models.
In the weakly coupled BCS theory, this value is $3.5$.
In the usual holographic superconductor model \cite{Hartnoll:2008vx,Hartnoll:2008kx,Horowitz:2010gk},
this value is approximately $8$, which is more than twice the one in the BCS theory,
but roughly approximates the value measured in high temperature superconductor materials \cite{Gomes:2007}.
The HD term introduced in holographic superconductor model drives the superconducting energy gap running,
ranging from $5.5$ to $16.2$ \cite{Wu:2010vr,Ma:2011zze,Mansoori:2016zbp,Wu:2017xki,Ling:2016lis}\footnote{The holographic superconductor models from HD theory coupling Weyl tensor
are also constructed in \cite{Momeni:2011ca,Momeni:2012ab,Momeni:2012uc,Roychowdhury:2012hp,Zhao:2012kp,Momeni:2013fma,Momeni:2014efa,Zhang:2015eea}.
But the study of conductivity is absent. Also, the extension of superconducting energy gap is also observed in other holographic superconductor models
from the Gauss-Bonnet gravity \cite{Gregory:2009fj,Pan:2009xa} and the quasi-topological gravity \cite{Kuang:2010jc,Kuang:2011dy}.
But the value of the superconducting energy gap is always greater than the value of the standard version holographic superconductor.}.
In this paper, we also study the properties of the conductivity of our present model and in particular the running of the superconducting energy gap.

Our paper is organized as what follows. We will introduce the framework of holographic superconductivity in Section~\ref{sec2}. In Section~\ref{sec-setup}, we make a deep analysis for the instabilities. Then we move on to investigate the condensation in Section~\ref{sec4}. In Section~\ref{sec5} we give the results of the conductivity.
Conclusion and discussion are given in Section~\ref{sec7}.

\section{Holographic framework}\label{sec2}

Our starting point is the following actions
\begin{subequations}
\label{action}
\begin{align}
\label{action-0}
S_0&=\frac{1}{2\kappa^2}\int d^4x\sqrt{-g}\Big(R+\frac{6}{L^2}\Big)
\,,
\
\\
\label{action-SPsi}
S_{\Psi}&=-\int d^4x\sqrt{-g}\Big(|D_{\mu}\Psi|^2+m^2|\Psi|^2-\alpha_1 L^2 C^2|\Psi|^2\Big)\,,
\\
\label{action-SA}
S_A&=-\int d^4x\sqrt{-g}\frac{L^2}{8g_F^2}F_{\mu\nu}X^{\mu\nu\rho\sigma}F_{\rho\sigma}\,.
\end{align}
\end{subequations}
$F=dA$ is the Maxwell field strength of gauge field $A$.
$D_{\mu}=\partial_{\mu}-iqA_{\mu}$ is the covariant derivative and $\Psi$ is the charged complex scalar field with mass $m$ and the charge $q$ of the Maxwell field $A$.
We can write $\Psi=\psi e^{i\theta}$ with $\psi$ being a real scalar field and $\theta$ a St\"uckelberg field.
And then, for convenience we choose the gauge $\theta=0$ in what follows.
In the action $S_{\Psi}$, a new interaction, which couples the complex scalar field to the Weyl tensor, is added in $S_{\Psi}$.
Since the pure AdS geometry is conformally flat,
the Weyl tensor vanishes in the UV boundary.
But $C^2$ provides a nontrivial source for the scalar in the bulk geometry
and gives an effective mass of the scalar.
This interaction provides a new mechanism of the symmetry breaking\footnote{This interaction for a neutral scalar field has been study in \cite{Gubser:2005ih,Myers:2016wsu,Lucas:2017dqa,Wu:2018xjy,Wu:2019vdu,Wu:2019orq}.},
for which we shall do in-depth studies in this paper.
In the action $S_A$, the tensor $X$ is
\fa
X_{\mu\nu}^{\ \ \rho\sigma}&=&
I_{\mu\nu}^{\ \ \rho\sigma}+\alpha_2|\Psi|^2I_{\mu\nu}^{\ \ \rho\sigma}-8\gamma_{1,1}L^2 C_{\mu\nu}^{\ \ \rho\sigma}
-4L^4\gamma_{2,1}C^2I_{\mu\nu}^{\ \ \rho\sigma}
-8L^4\gamma_{2,2}C_{\mu\nu}^{\ \ \alpha\beta}C_{\alpha\beta}^{\ \ \rho\sigma}
\nonumber
\\
&&
-4L^6\gamma_{3,1}C^3I_{\mu\nu}^{\ \ \rho\sigma}
-8L^6\gamma_{3,2}C^2C_{\mu\nu}^{\ \ \rho\sigma}
-8L^6\gamma_{3,3}C_{\mu\nu}^{\ \ \alpha_1\beta_1}C_{\alpha_1\beta_1}^{\ \ \ \alpha_2\beta_2}C_{\alpha_2\beta_2}^{\ \ \ \rho\sigma}
+\ldots
\,.
\label{X-tensor}
\ffa
$I_{\mu\nu}^{\ \ \rho\sigma}=\delta_{\mu}^{\ \rho}\delta_{\nu}^{\ \sigma}-\delta_{\mu}^{\ \sigma}\delta_{\nu}^{\ \rho}$
is an identity matrix and $C^n=C_{\mu\nu}^{\ \ \alpha_1\beta_1}C_{\alpha_1\beta_1}^{\ \ \ \alpha_2\beta_2}\ldots C_{\alpha_{n-1}\beta_{n-1}}^{\ \ \ \mu\nu}$
with $C_{\mu\nu\rho\sigma}$ being the Weyl tensor.
When $X_{\mu\nu}^{\ \ \rho\sigma}=I_{\mu\nu}^{\ \ \rho\sigma}$, the action $S_A$ reduces to the standard Maxwell theory.
The second term introduces the interaction between the scalar field and the gauge field.
Starting from the third term, they are an infinite family of HD terms \cite{Witczak-Krempa:2013aea}.
In this paper, we mainly focus on the top four terms in $X_{\mu\nu}^{\ \ \rho\sigma}$.
For convenience sake, we denote $\gamma_{1,1}=\gamma$ and $\gamma_{2,i}=\gamma_i (i=1,2)$.
In SS-AdS black brane background, when other parameters are turned off,
$\gamma$ and $\gamma_1$ are constrained in the region $-1/12\leq\gamma\leq 1/12$ \cite{Myers:2010pk,Ritz:2008kh} and $\gamma_1\leq 1/48$ \cite{Witczak-Krempa:2013aea}, respectively.
These constraints come from the instabilities and causality of the vector modes.
However, for the black brane with scalar hair, we must reexamine the instabilities and causality of the vector modes and we leave them for future.
In this paper, we shall constraint these coupling parameters in small region, which is safe.

\section{Superconducting instability}\label{sec-setup}

When $\Psi=0$, the system \eqref{action} achieves a charged black brane solution, which corresponds to the normal phase.
In this section, we shall explore the instability condition for the normal phase towards the development of a hairy black brane under small charged scalar field perturbations.
This allows one to determine the superconducting phase structure in the dual boundary field theory.
The condition for the formation of the hairy black brane depends on the charge and the mass of the scalar field
as well as the background which is specified by the model parameters.
For clarity, we first study the role of $\alpha_1$ term in the formation of the hairy black brane.
And then, we further explore the joint effect from the $\alpha_1$ term and the $\gamma$ term.

\subsection{Case I: $\alpha_1$ HD term}

In this subsection, we first want to see what role the $\alpha_1$ term plays in the formation of the hairy black brane.
So we only turn on $\alpha_1$ term and turn off other coupling parameters in this section.
In this case, the background geometry of the normal phase is RN-AdS black brane,
\fa
\label{bl-br-r}
&&
ds^2=\frac{1}{u^2}\Big(-f(u)dt^2+dx^2+dy^2\Big)+\frac{1}{u^2f(u)}du^2\,,
\nonumber
\\
&&
f(u)=(1-u)p(u)\,,~~~~~~~
p(u)=1+u+u^2-\frac{\mu^2u^3}{4}\,,
\nonumber
\\
&&
A_t(u)=\mu(1-u)\,.
\ffa
Here $\mu$ is the chemical potential of the field theory.
The Hawking temperature is
\f
T=\frac{p(1)}{4\pi}=\frac{1}{4\pi}\Big(\frac{\mu^2}{4}-3\Big).
\ff

\begin{figure}
\center{
\includegraphics[scale=0.65]{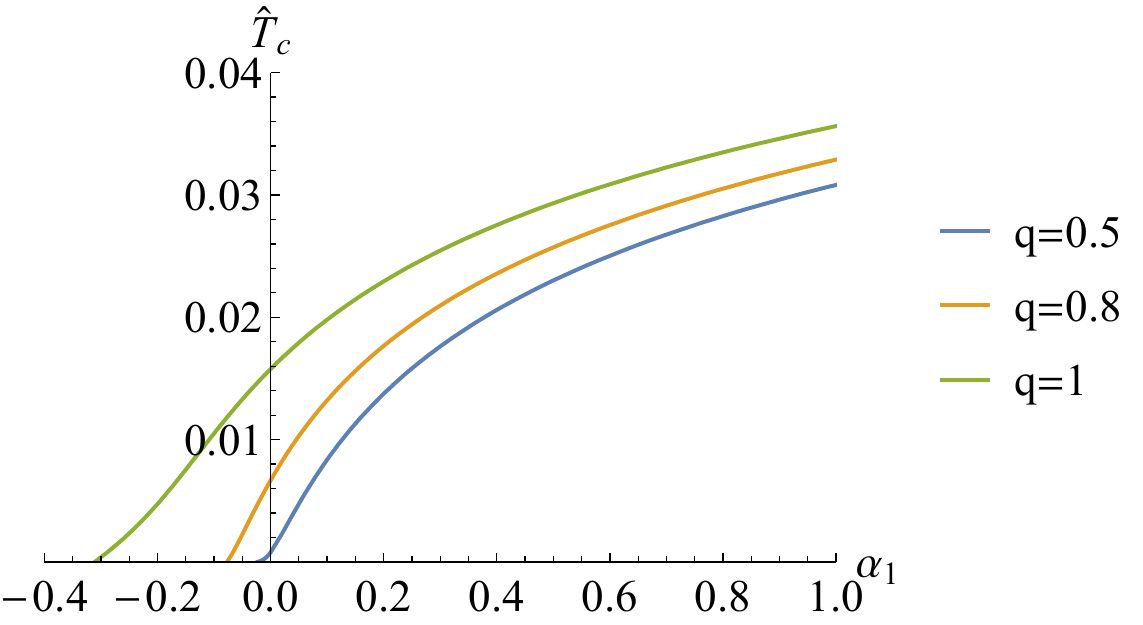}\ \hspace{0.05cm}
\includegraphics[scale=0.55]{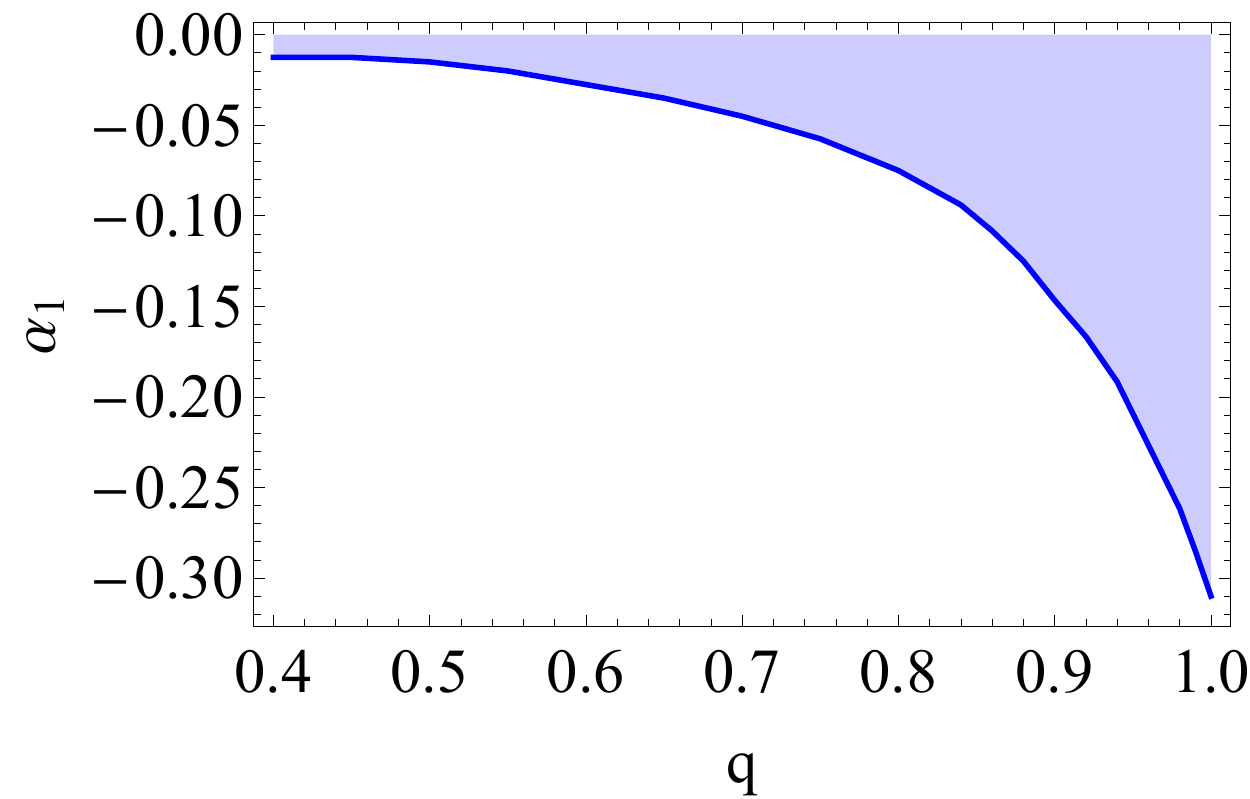}\ \hspace{0.05cm}
\caption{\label{fig-Tcvsalpha1}Left plot: Phase diagram $(\alpha_1,\hat{T}_c)$ with $m^2=-2$ for different $q$.
Right plot: QPT diagram in $(\alpha_1,q)$ space for $m^2=-2$. The blue line is the QPT critical line and the blue zone is the superconducting phase at zero temperature.
}}
\end{figure}

We can estimate the critical temperature for the formation of the superconducting phase
by finding static normalizable modes of the scalar field $\psi$ in the above background \eqref{bl-br-r}.
The procedure has been used in \cite{Horowitz:2013jaa,Ling:2014laa,Ling:2017naw,Kim:2015dna}.
This problem can be casted into a positive self-adjoint eigenvalue problem for $q^2$ and so we write the equation of motion for the scalar field as the following form:
\begin{eqnarray}
\label{eom-psi-eigen}
\big[\nabla^2-m^2+\alpha_1L^2C^2\big]\psi=q^2A^2\psi\,.
\end{eqnarray}
Without loss of generality, we set the mass of the charged scalar field as $m^2=-2$ here\footnote{We have set $L=1$.}.
For this case, its asymptotical behavior at infinity is
\begin{eqnarray}
\label{psi-asy}
\psi=\psi_1 u+\psi_2 u^2\,.
\end{eqnarray}
Here, we will treat $\psi_1$ as the source and $\psi_2$ as the expectation value in the dual boundary theory.
Also we set $\psi_1=0$ such that the condensation is not sourced.
Now, this system is determined by the scaling-invariant Hawking temperature $\hat{T}=T/\mu$, the charge of complex scalar field $q$ and the coupling parameter $\alpha$.
When the non-trivial scalar profile is developed, the superconducting phase forms.
Therefore, we can numerically solve the above equation \eqref{eom-psi-eigen} to locate the critical temperature of the superconducting phase
as the function of $\alpha_1$ for different $q$, which is shown in the left plot in FIG.\ref{fig-Tcvsalpha1}.
Note that the region above the line is the normal state and the region below the line is the superconducting phase.
This plot exhibits that when we reduce the temperature for given parameter $q$ and $\alpha$, an thermodynamic phase transition happens.
In addition, at zero temperature, we find that for given $q$ when the coupling parameter $\alpha_1$ increases,
the superconducting phase also appears, which is a quantum phase transition (QPT).
The phase diagram $(\alpha_1,q)$ is shown in the right plot in FIG.\ref{fig-Tcvsalpha1}.
There are several characteristics are summarized as what follows:
\begin{itemize}
  \item For a given $\alpha_1$, we find that the critical temperature becomes higher with the increase of the charge (left plot in FIG.\ref{fig-Tcvsalpha1}),
  which means that the increase of the charge make the condensation easier.
  This tendency is consistent with our intuition and have been observed in \cite{Horowitz:2013jaa,Ling:2014laa}.
  \item For a given charge, we find a rise in critical temperature with the coupling parameter $\alpha_1$ (left plot in FIG.\ref{fig-Tcvsalpha1}),
  which means that the higher derivative term $\alpha_1$ plays the role of driving the superconducting phase transition.
  \item From the left plot in FIG.\ref{fig-Tcvsalpha1}, we see that for a given charge, if the coupling parameter $\alpha_1$ is relatively small,
  the system would not undergo a superconducting phase transition no matter how low the temperature.
  It means that there is a QPT at zero temperature. We show the QPT diagram in $(\alpha_1,q)$ space in the right plot in FIG.\ref{fig-Tcvsalpha1}.
  The blue line is the QPT critical line and the blue zone is the superconducting phase at zero temperature.
  The QPT can be also understood by the BF bound.
  We shall further address this problem in what follows.
\end{itemize}

Now, we shall analyze the superconducting phase transition at zero temperature by BF bound.
To have a superconducting phase transition, the near horizon effective mass squared shall be below the corresponding BF bound,
but the UV boundary effective mass squared shall be above the boundary AdS$_4$ BF bound, which guarantees the stability of scalar field at the boundary.

For the RN-AdS black brane, the extremal limit can be arrived at when $\mu=2\sqrt{3}$.
At this limit, the near horizon geometry and the gauge field are \cite{Faulkner:2009wj}
\begin{subequations}
\label{geometry-nh}
\begin{align}
\label{ds-nh}
ds^2&=\frac{L_2^2}{\zeta^2}(-d\tau^2+d\zeta^2)+dx^2+dy^2\,,
\
\\
\label{Atau-nh}
A_{\tau}&=\frac{\mu L_2^2}{\zeta}\,,
\end{align}
\end{subequations}
where $L_2\equiv L/\sqrt{6}$.
In deriving the above equations, we have made the transformation $(1-u)=\epsilon\frac{L_2^2}{\zeta}$ and $t=\epsilon^{-1}\tau$
in the limit $\epsilon\rightarrow 0$ with finite $\zeta$ and $\tau$.
It is obvious that at the zero-temperature limit, the near horizon geometry is
$\rm AdS_2\times\mathbb{R}^2$ with the $\rm AdS_2$ curvature radius $L_2$.

The effective mass of this system is
\fa
m_{eff}^2=m^2+q^2A^2-\alpha_1L^2C^2\,.
\label{meff}
\ffa
Since $q^2A^2=q^2g^{tt}A_t^2$, which contributes with minus sign,
it tends to destabilize the normal state and induces the superconductivity.
While the role plays $\alpha_1$ term depends on the sign of $\alpha_1$,
which can stabilize or destabilize the normal state.
To have a superconducting phase transition, we require that
the near horizon effective mass squared shall be below the corresponding BF bound,
but the UV boundary effective mass squared shall be above the corresponding BF bound.
Thus we have
\begin{align}
\label{BFcondition1}
m_{IR}^2<-\frac{1}{4L_2^2}\,,
~~~~~~
m_{UV}^2\geq-\frac{9}{4}\,,
\end{align}
where $m_{IR}^2$ and $m_{UV}^2$ denote the near horizon effective mass squared and the bulk mass squared, respectively.
Using the expression of the effective mass squared \eqref{meff} combining with the near horizon geometry \eqref{geometry-nh} and the AdS$_4$ geometry,
the above equations can be specifically expressed as
\begin{align}
\label{BFcondition2}
m^2-2q^2-8\alpha_1<-\frac{3}{2}\,,
~~~~~~
m^2\geq-\frac{9}{4}\,.
\end{align}

\begin{figure}
\center{
\includegraphics[scale=0.5]{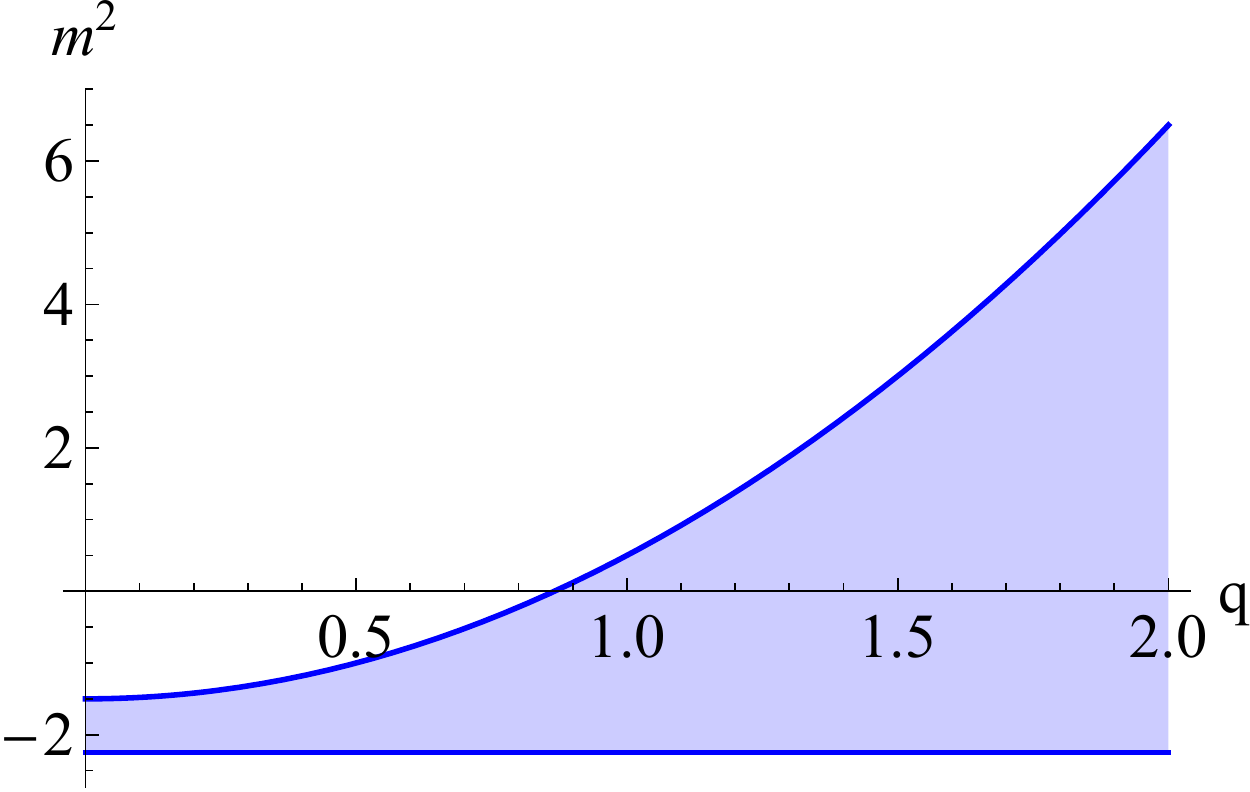}\ \hspace{0.5cm}
\includegraphics[scale=0.5]{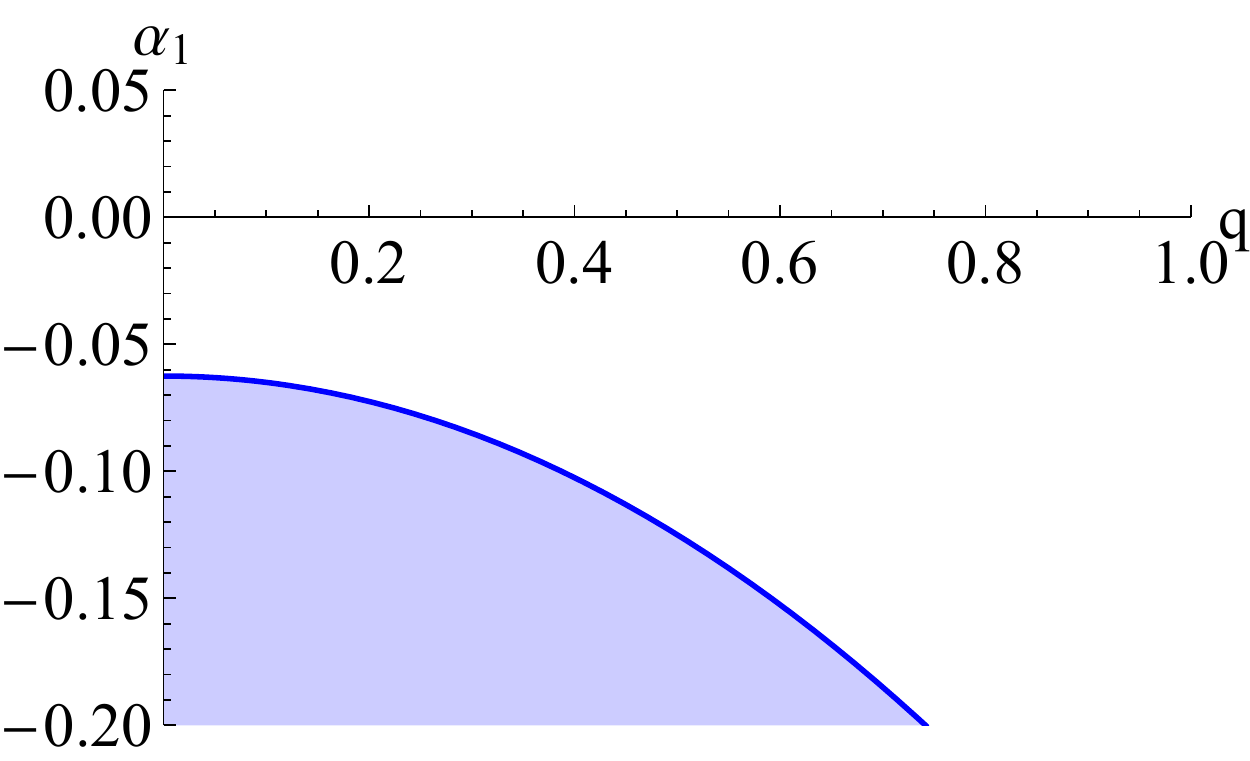}\ \hspace{0.5cm}
\caption{\label{fig-m2vsq-alpha2vs0} Left plot: The blue zone is the allowed region for the superconducting phase transition at $m^2$ vs $q$ plane at the zero temperature for $\alpha_1=0$.
Right plot: The blue zone is the allowed region for the superconducting phase transition at the zero temperature for $m^2=-2$.}}
\end{figure}

It has been well studied that for the usual holographic superconductor ($\alpha_1=0$ here),
the superconducting instabilities are determined by both the bulk mass squared $m^2$ and the charge $q$, equivalently, the chemical potential,
which can be clearly seen from the above instability condition \eqref{BFcondition2}
and have been explored in \cite{Denef:2009tp,Kim:2009kb}.
Here, we also exhibit the allowed region (blue zone) for the superconducting phase transition with $\alpha_1=0$ at $m^2$ vs $q$ plane at the zero temperature (left plot in FIG.\ref{fig-m2vsq-alpha2vs0}).
For small $q$, the superconducting phase transition is forbade if the mass squared is positive.
But provided the charge is large, the superconducting phase transition still happen for $m^2=0$ or even positive $m^2$,
which is called the density driven symmetry breaking in holographic superconductor in \cite{Kim:2009kb}.

Subsequently, we turn to explore what role $\alpha_1$ plays.
We plot the $\alpha_1$ as the function of $q$ (i.e., QPT diagram) for $m^2=-2$ at zero temperature,
which is shown in the right plot in FIG.\ref{fig-m2vsq-alpha2vs0}.
The result is qualitatively consistent with that above by finding static normalizable modes of the scalar field
(see the right plot in FIG.\ref{fig-Tcvsalpha1}).

\begin{figure}
\center{
\includegraphics[scale=0.45]{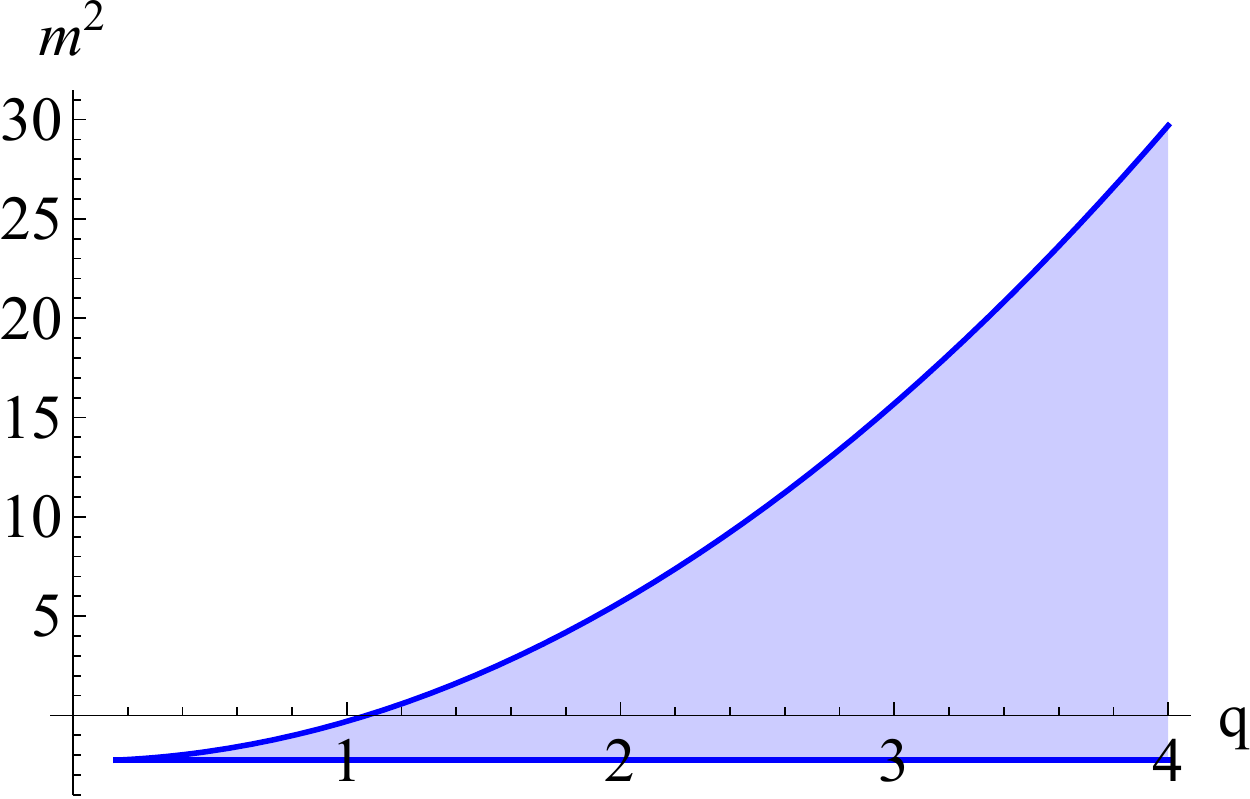}\ \hspace{0.5cm}
\includegraphics[scale=0.45]{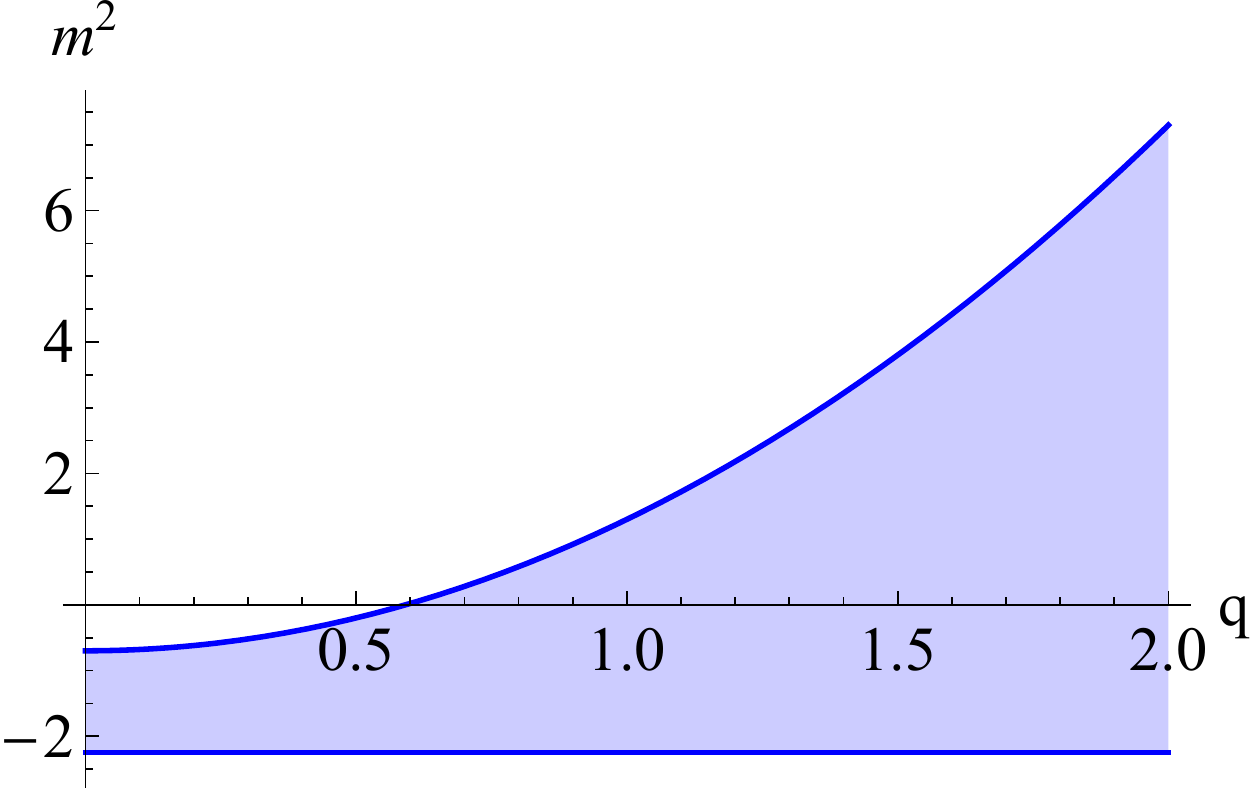}\ \hspace{0.5cm}
\caption{\label{fig-m2vsq} The blue zone is the allowed region for the superconducting phase transition at $m^2$ vs $q$ plane at the zero temperature (left plot is for $\alpha_1=-0.1$ and right plot is for $\alpha_1=0.1$).}}
\end{figure}
\begin{figure}
\center{
\includegraphics[scale=0.55]{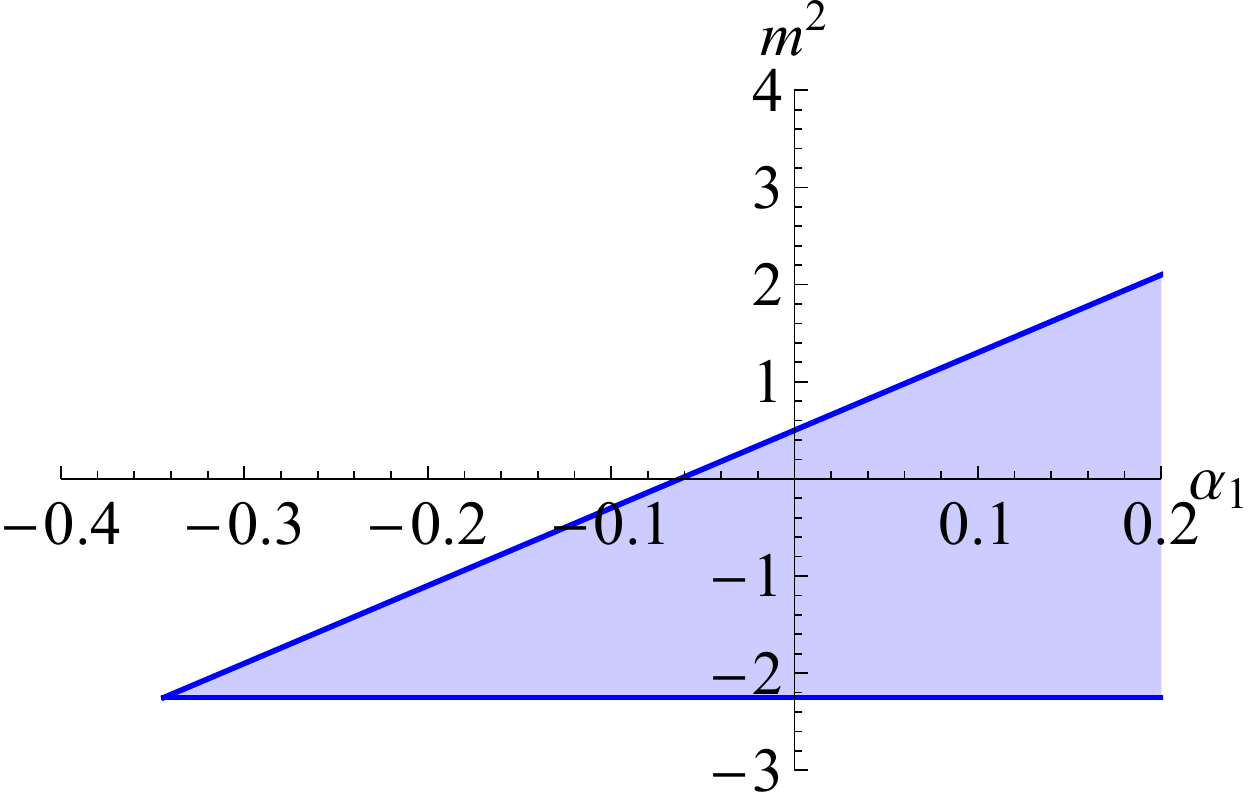}\ \hspace{0.2cm}
\includegraphics[scale=0.55]{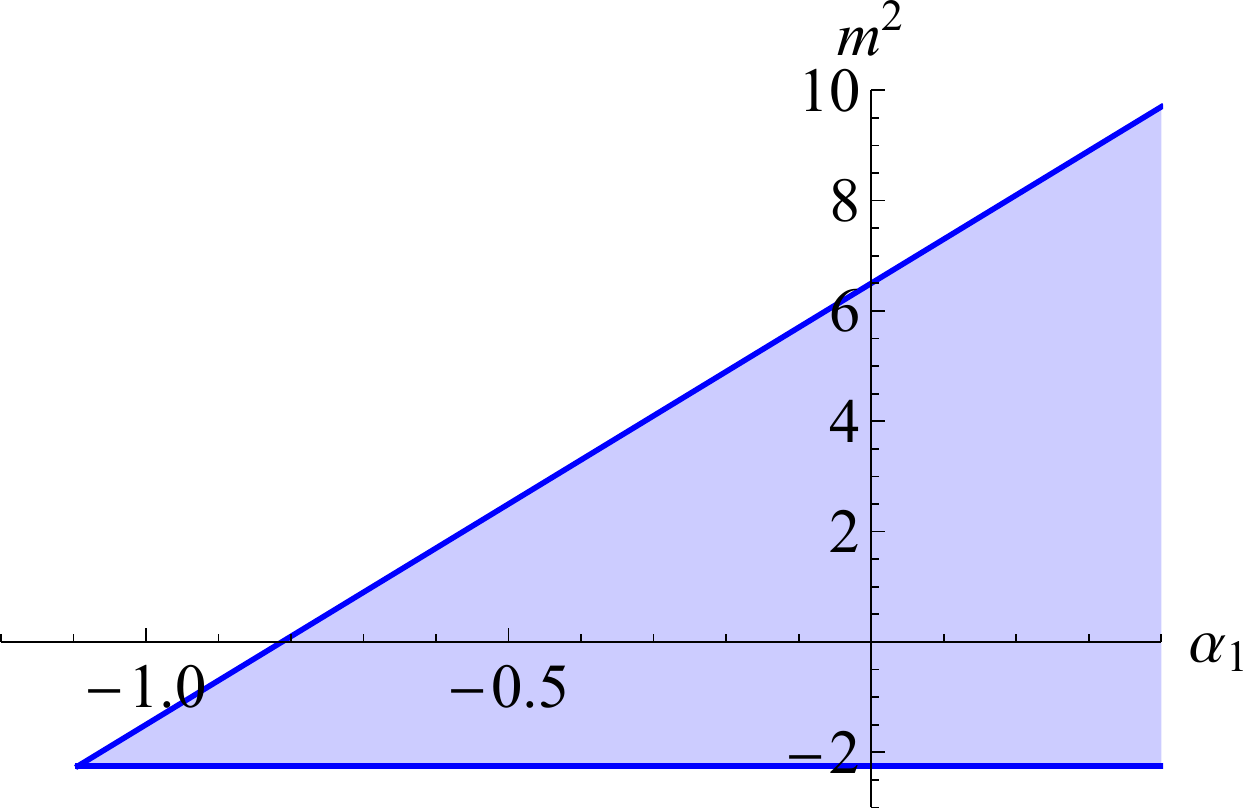}\ \hspace{0.2cm}
\caption{\label{fit_q} The blue zone is the allowed region for the superconducting phase transition at $m^2$ vs $\alpha_1$ plane at the zero temperature (left plot is for $q=1$ and right plot is for $q=2$).}}
\end{figure}
\begin{figure}
\center{
\includegraphics[scale=0.6]{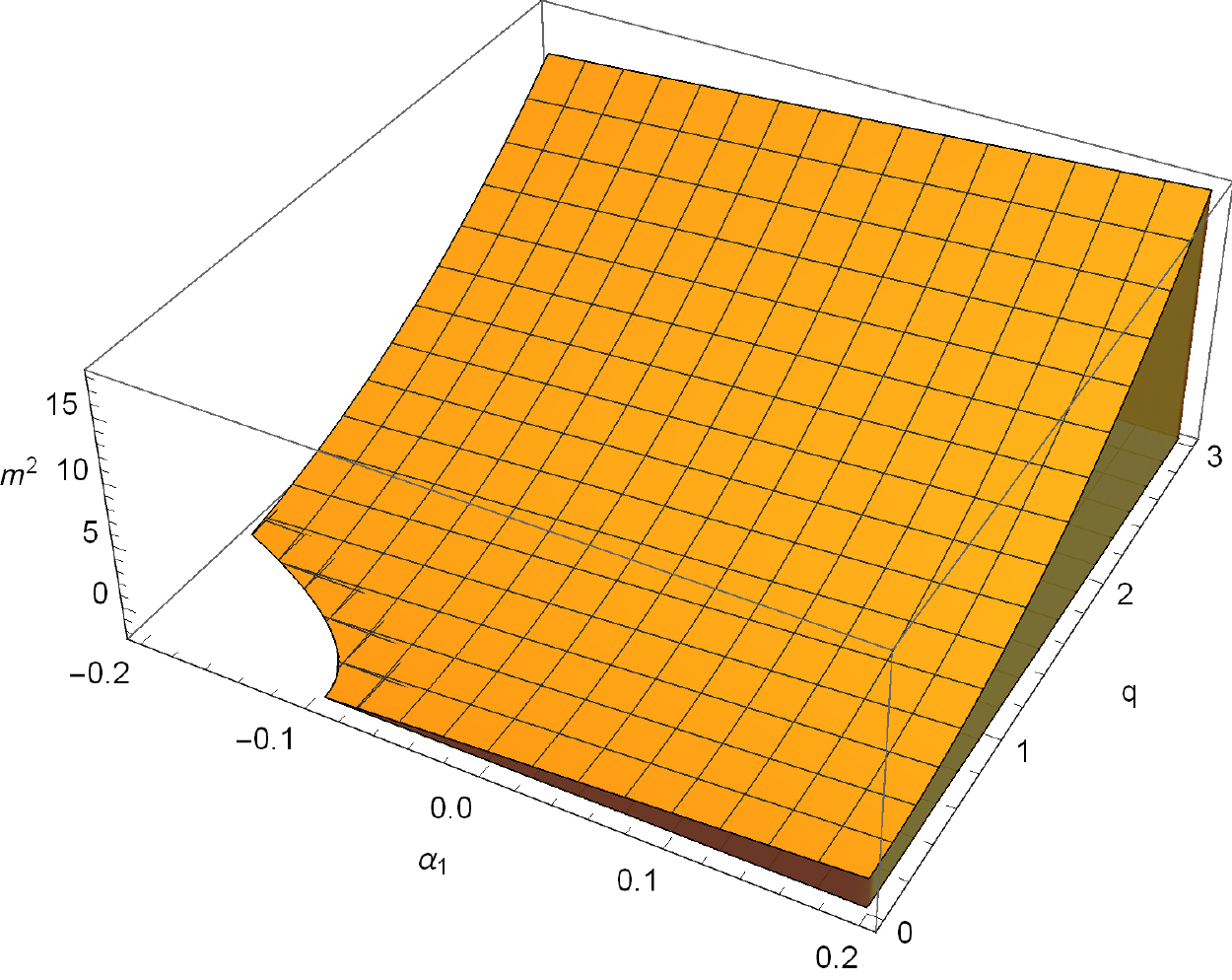}\\
\caption{\label{fit_gamma} 3D plot of the allowed region for the superconducting phase transition at the zero temperature.}}
\end{figure}

Further, we plot the allowed region for the superconducting phase transition at $m^2$ vs $q$ plane
at the zero temperature for $\alpha_1=-0.1$ and $\alpha_1=0.1$, respectively (FIG.\ref{fig-m2vsq}).
It is obvious that for $\alpha_1=-0.1$, if $q$ is less than some critical value ($q\simeq 0.158$),
the superconducting phase transition is forbade regardless of the value of $m^2$ (see the left plot in FIG.\ref{fig-m2vsq}).
For $\alpha_1=0.1$, the allowed range of $m^2$ for the superconducting phase transition becomes larger than $\alpha_1=0$ for the fixed $q$.
To address this point more clearly, we also plot the relation between $m^{2}$ and $\alpha_{1}$ for $q=1$ and $q=2$ in FIG.\ref{fit_q}
and a 3D plot for $q$, $m^2$ and $\alpha_1$ in FIG.\ref{fit_gamma}.
It indicates that one can tune $\alpha_1$ to trigger a quantum phase transition.
Also, we can infer that for small $q$, the superconducting phase transition still happen for $m^2=0$ or even positive $m^2$,
provided that $\alpha_1$ is large. It is just the role $\alpha_1$ plays and we call the HD driven symmetry breaking in holographic superconductor,
which is the focus of our present paper and we shall thoroughly explore this issue in next section.


\subsection{Case II: $\gamma$ HD term}\label{sec-gamma}

In this subsection, we simultaneously turn on $\alpha_1$ and $\gamma$.
Since $\gamma$ term involves the coupling between the Weyl tensor and gauge field,
the background geometry of the normal phase is no longer a RN-AdS black brane.
We need to solve a set of third order differential equations to obtain the background solution.
It is a hard task.
However, we can obtain the perturbative solution up to the first order of $\gamma$ as \cite{Ling:2016dck}
\fa
\label{b2-br-r}
&&
ds^2=\frac{f(u)}{u^{2}}dt^{2}+\frac{1}{u^{2}f(u)}du^{2}+[\frac{1}{u^{2}}(1+\gamma \frac{u^{2}\mu^{2}}{9})](dx^{2}+dy^{2})\,,
\nonumber
\\
&&
f(u)=(1-u)p(u)\,,
\nonumber
\\
&&
p(u)=1+u+u^2-\frac{\mu^2u^3}{4}
+\frac{1}{180}u^{3}\gamma \mu^{2}[240-28\mu^{2}+26u^{4}\mu^{2}+(u+u^{2}+u^{3})(-100+\mu^{2})]\,,
\nonumber
\\
&&
A_t(u)=\frac{1}{90}\mu \{90+74u^{5}\gamma \mu^{2}+45u^{4}\gamma (-4-\mu^{2})-[90+\gamma (180+29 \mu^{2})]\}\,.
\ffa
For this black brane, we can obtain the dimensionless Hawking temperature $\hat{T}\equiv T/\mu$ with
\begin{equation}\label{T}
T=\frac{12-\mu^{2}}{16\pi \mu}+\gamma \frac{\mu(\mu^{2}-60)}{720\pi}\,.
\end{equation}
At the zero temperature limit, the chemical potential $\mu$ becomes
\begin{equation}\label{cp}
\mu=\sqrt{\frac{3}{2}}\sqrt{20+\frac{15}{\gamma}-\frac{\sqrt{5}\sqrt{45+72\gamma+80\gamma^{2}}}{\gamma}}.
\end{equation}
When $\gamma\rightarrow 0$, $\mu=2\sqrt{3}$, which reduces the case of RN-AdS black brane.
At extremal limit, the near horizon geometry of the perturbative solution \eqref{b2-br-r} is also $\rm AdS_2\times\mathbb{R}^2$ as RN-AdS black brane but
with a different $\rm AdS_2$ curvature radius $L_2$ as
\fa
\label{L22}
L_2^2=
\frac{-4000 \gamma ^2+5 \sqrt{5} (40 \gamma +33) \sqrt{80 \gamma ^2+72 \gamma +45}-4956
   \gamma -2475}{4 \gamma }\,,
\ffa
which explicitly dependens on the Weyl parameter $\gamma$.

Following the same procedure in the above subsection, in this case,
the constraint on the the near horizon effective mass squared gives
\begin{eqnarray}
&&
m^{2}+\frac{4\alpha_{1}}{3}-\frac{2475+4956\gamma+4000\gamma^{2}-165w-200w}{16\gamma}+
\nonumber
\\
&&
\frac{q^{2}(2355+2860\gamma-161w)^{2}(-15-20\gamma+w)}{600[-4000\gamma^{2}+165(-15+w)+4\gamma(-1239+50w)]}<-\frac{3}{2}\,.
\label{BFcondition3}
\end{eqnarray}
For convenience, in the above equation, we have defined $w$ as
\begin{equation}\label{2}
w=\sqrt{5}\sqrt{45+72\gamma+80\gamma^{2}}\,.
\end{equation}
In addition, the bulk mass squared satisfies
\begin{eqnarray}
m^{2}>-\frac{9}{4}\,.
\label{BFcondition4}
\end{eqnarray}
Eqs.\eqref{BFcondition3} and \eqref{BFcondition4} give the conditions that the superconducting phase happens at zero temperature.
\begin{figure}
\center{
\includegraphics[scale=0.45]{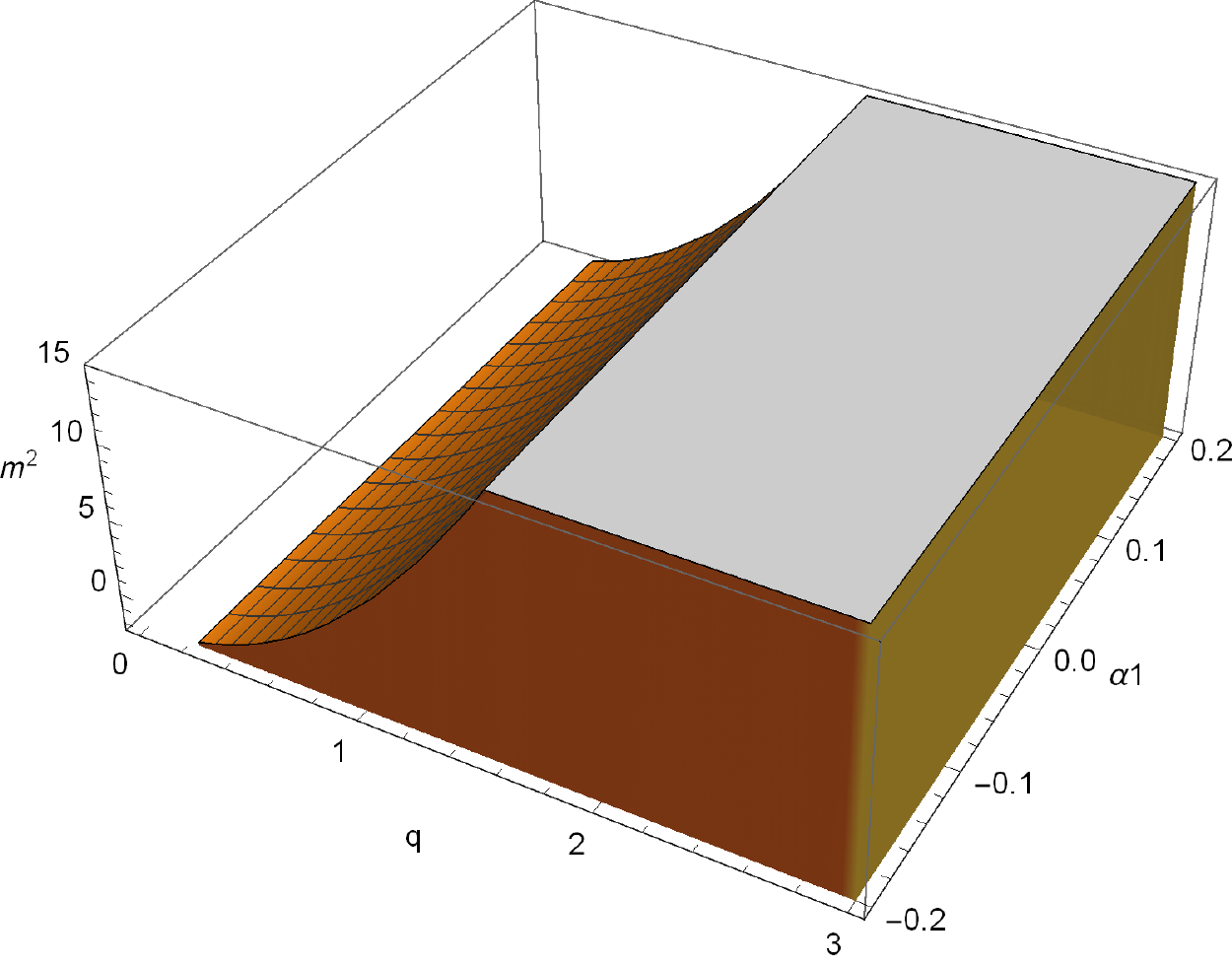}\ \hspace{0.8cm}
\includegraphics[scale=0.45]{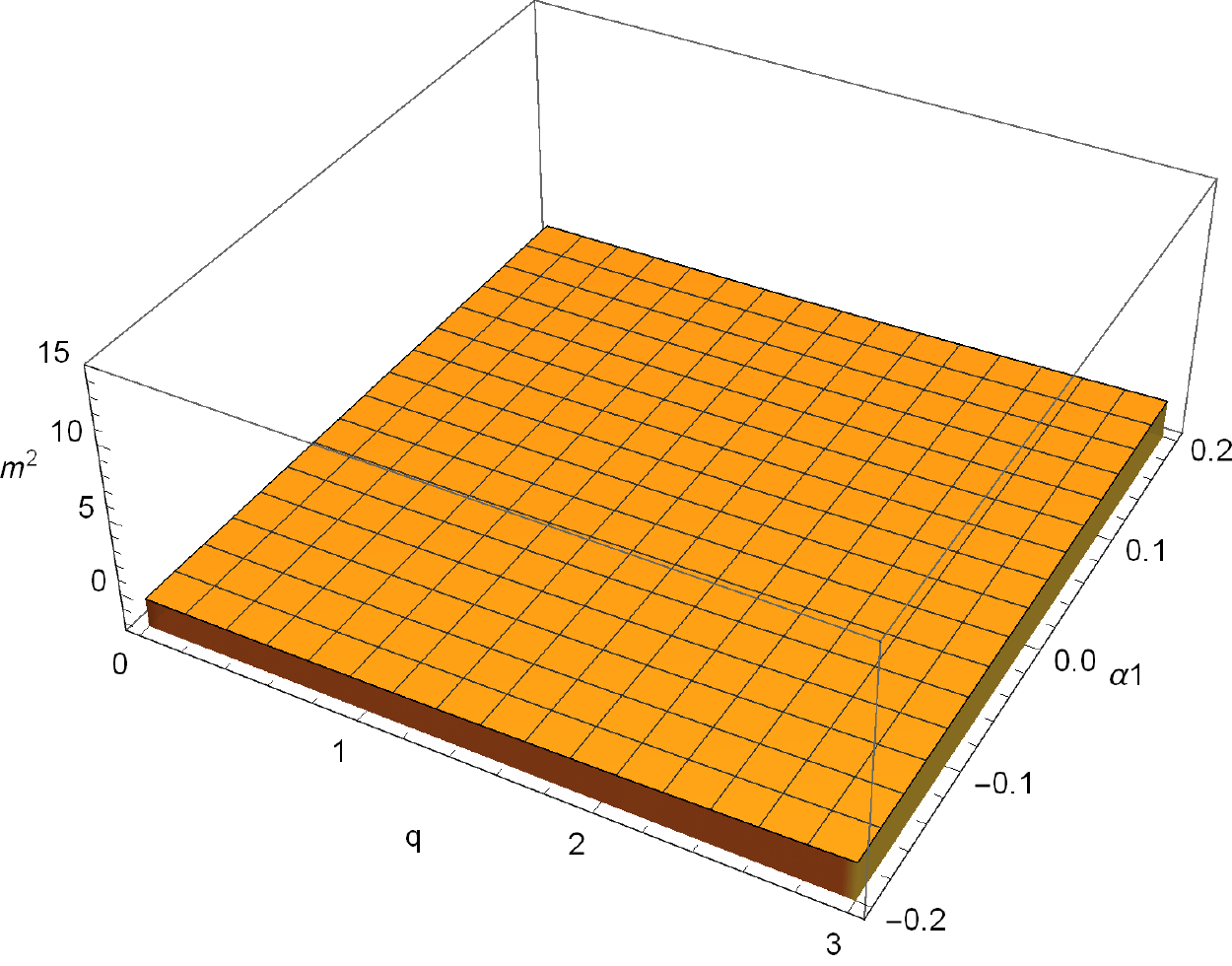}\ \hspace{0.2cm}
\caption{\label{fit_gamma3D} 3D plot of the allowed region for the superconducting phase transition at the zero temperature for a fixed $\gamma$ (left plot is for $\gamma=-1/12$ and right plot is for $\gamma=1/12$).}}
\end{figure}
\begin{figure}
\center{
\includegraphics[scale=0.4]{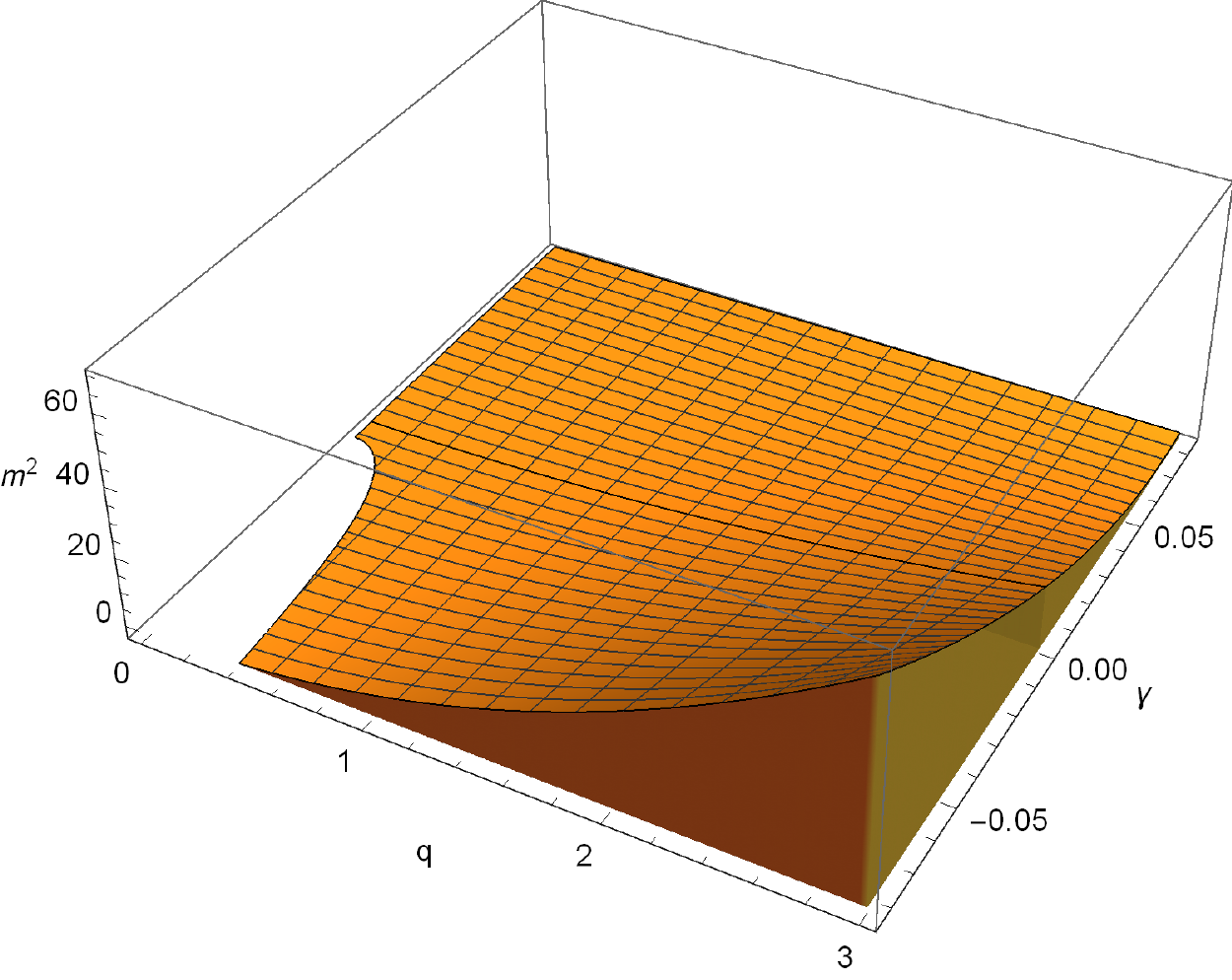}\ \hspace{0.01cm}
\includegraphics[scale=0.4]{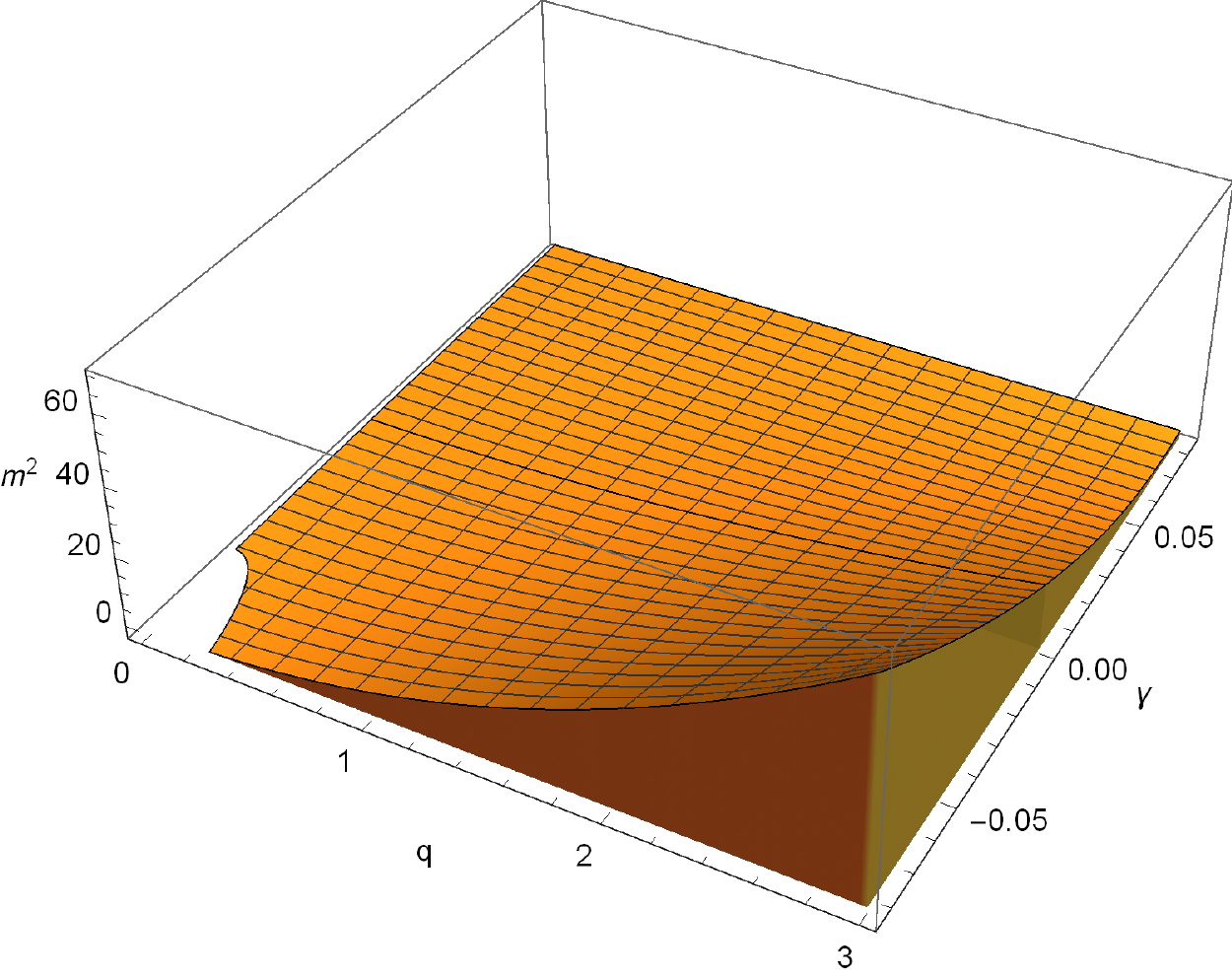}\ \hspace{0.01cm}
\includegraphics[scale=0.4]{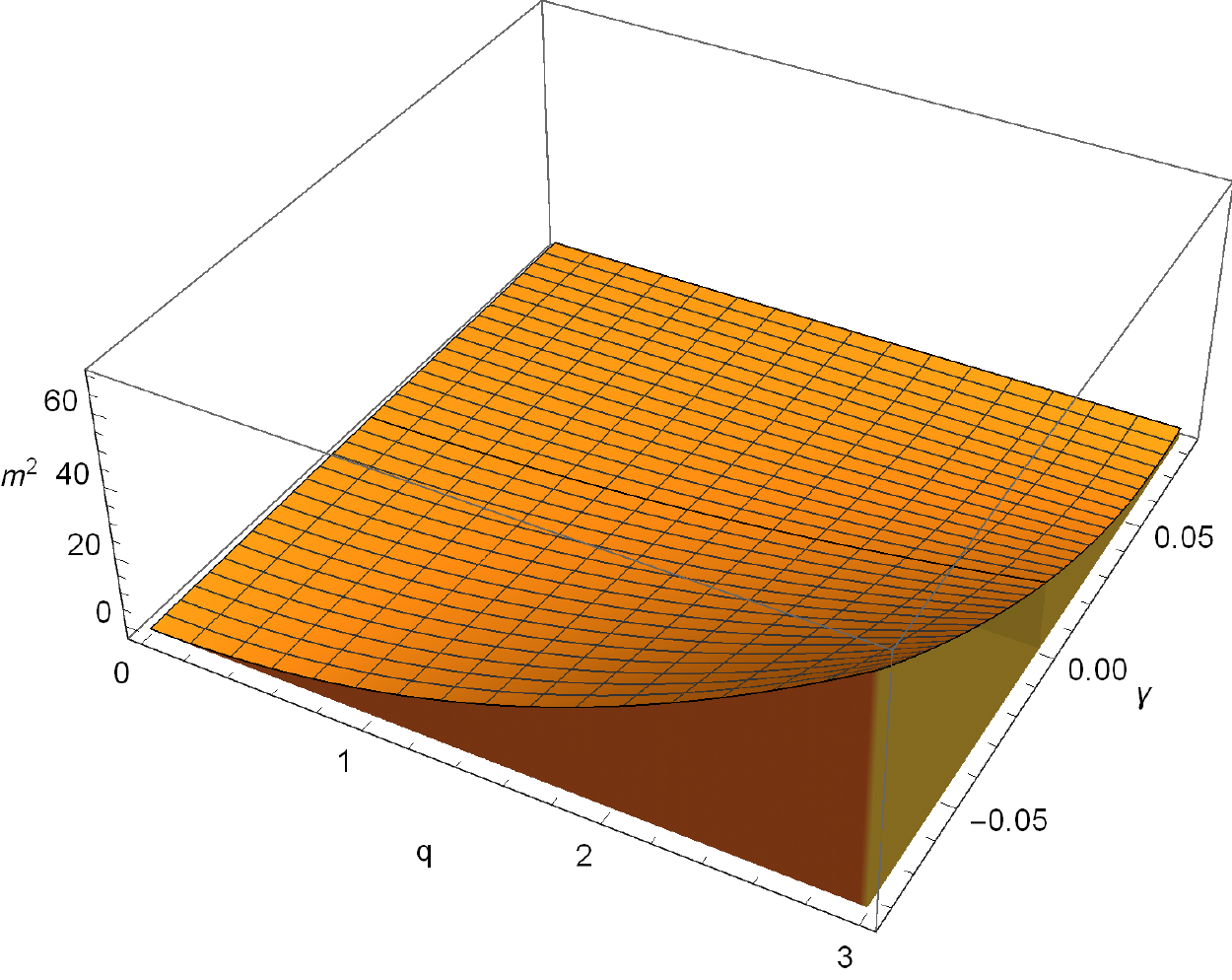}\ \hspace{0.01cm}
\caption{\label{fit_alphaf} 3D plot of the allowed region for the superconducting phase transition at the zero temperature for a fixed $\alpha_1$ (from left to right, $\alpha_1=-0.5, 0, 0.5$).}}
\end{figure}
We summarize the roles $\alpha_1$ and $\gamma$ play in the superconducting phase transition as what follows.
\begin{itemize}
  \item In previous subsection, we have observed that for $\gamma=0$, in the phase diagram $(m^2,q,\alpha_1)$,
  there is a region in which the superconducting phase transition is forbade.
  When $\gamma=-1/12$, such a region still holds (left plot in FIG.\ref{fit_gamma3D}).
  But when $\gamma=1/12$, such a forbidden region of superconducting phase transition vanishes (right plot in FIG.\ref{fit_gamma3D}).
  It indicates that we can find a parameter space, the superconducting phase transition can always happen.
  \item FIG.\ref{fit_alphaf} exhibits 3D plot of the allowed region for the superconducting phase transition at the zero temperature for a fixed $\alpha_1$.
  When $\alpha_1$ is negative, there is a forbidden region of superconducting phase transition in the phase diagram $(m^2,q,\gamma)$.
  As $\alpha_1$ increases, this region shrinks and $\alpha_1$ is large, this region vanishes.
\end{itemize}

In this section, we have made the superconducting instability analysis,
from which we clearly see what roles of the HD terms $\alpha_1$ and $\gamma$ play in the superconducting phase transition.
However, we would like to point out that the instability analysis is implemented in the probe limit
and it only provide a clue of the phase transition. To further confirm the phase transition, we need numerically solve the system \eqref{action}.

\section{Condensation}\label{sec4}

In this section, we shall numerically solve the system \eqref{action} to study the superconducting phase transition.
However, it is hard to solve the system \eqref{action} with backreaction
because it involves solving a set of third order differential equations
with high nonlinearity.
Therefore, we shall work in the probe limit, i.e.,
we don't consider the backreaction of the gauge field and the scalar field on the geometry.

In the probe limit, we consider the metric as
\fa
\label{bl-br}
&&
ds^2=\frac{1}{u^2}\Big(-f(u)dt^2+dx^2+dy^2\Big)+\frac{L^2}{u^2f(u)}du^2\,,
\nonumber
\\
&&
f(u)=(1-u)p(u)\,,~~~~~~~
p(u)=u^2+u+1\,,
\ffa
which is the SS-AdS black brane.
$u=1$ denotes the horizon and $u=0$ is the asymptotically AdS boundary.
The Hawking temperature of this system is
$
T=3/4\pi
$.
And then, the EOMs of gauge field and scalar field can be derived as
\begin{eqnarray}
\label{eom-gauge-psi}
\nabla_{\nu}\big(X^{\mu\nu\rho\sigma}F_{\rho\sigma}\big)
-4q^2A^{\mu}\psi^2
=0\,,
\,\,\,\,\,\,\,\,\,\,\,\,\,
\big[\nabla^2-(m^2+q^2A^2)+\alpha_1 C^2\big]\psi=0\,.
\end{eqnarray}
The ansatz for the scalar field and gauge field is taken as
\fa
\psi=\psi(u)\,,\,\,\,\,\,
A=\phi(u) dt\,.
\label{At}
\ffa
Under the above ansatz (Eqs.\eqref{bl-br} and \eqref{At}),
the EOMs for $\psi$ and $\phi$ (Eq.\eqref{eom-gauge-psi}) can be explicitly expressed as follows,
\begin{subequations}
\label{eoms-psi-phi}
\begin{align}
&
\label{psi}
\psi (u) \left(-\frac{q^2 \phi (u)^2}{u^3-1}+12 \alpha _1 u^4-u\right)-\left(u^3-1\right)
   \psi ''(u)-3 u^2 \psi '(u)=0\,,
\
\\
&
\label{phi}
(u^3-1)(48u^6 \gamma_{1}+8u^3 \gamma-1)\phi ''(u)+24u^{2}(u^{3}-1)(\gamma+12u^{3}\gamma_{1})\phi '(u)-2q^{2}\psi(u)^{2}\phi(u)=0.
\end{align}
\end{subequations}
Here, we only consider the HD terms up to $6$ order.
Regardless of any details of the above EOMs, the asymptotical behaviors of $\phi$ and $\psi$ at the conformal boundary are
\fa\label{bc0}
\phi=\mu-\rho u\,,\qquad\psi=\psi_1 u^{3-\Delta}+\psi_2 u^{\Delta}\,.
\ffa
In the dual boundary field theory, $\mu$ and $\rho$ are the chemical potential and the charge density, respectively, as has been mentioned above.
This system is depicted by a dimensionless quantity $\hat{T}\equiv T/\mu$.
We take standard quantization and so $\psi_1$ is treated as the source and $\psi_2$ as the expectation value of the scalar operator, for which we denote $\langle O_2\rangle$.
We expect that the condensation is not sourced and so we set $\psi_1=0$.
Subsequently, we numerically solve EOMs \eqref{eoms-psi-phi} by the shooting method
and study the properties of the condensation $\langle O_2\rangle$ with HD derivative terms\footnote{In this paper,
we only focus on the cases of $\alpha_1$ and $\gamma$ HD terms.}.
We shall firstly study the condensation by considering the scalar field with negative mass, $m^2=-2$,
which has been well studied in the usual holographic superconductor model.

\begin{figure}
\center{
\includegraphics[scale=0.8]{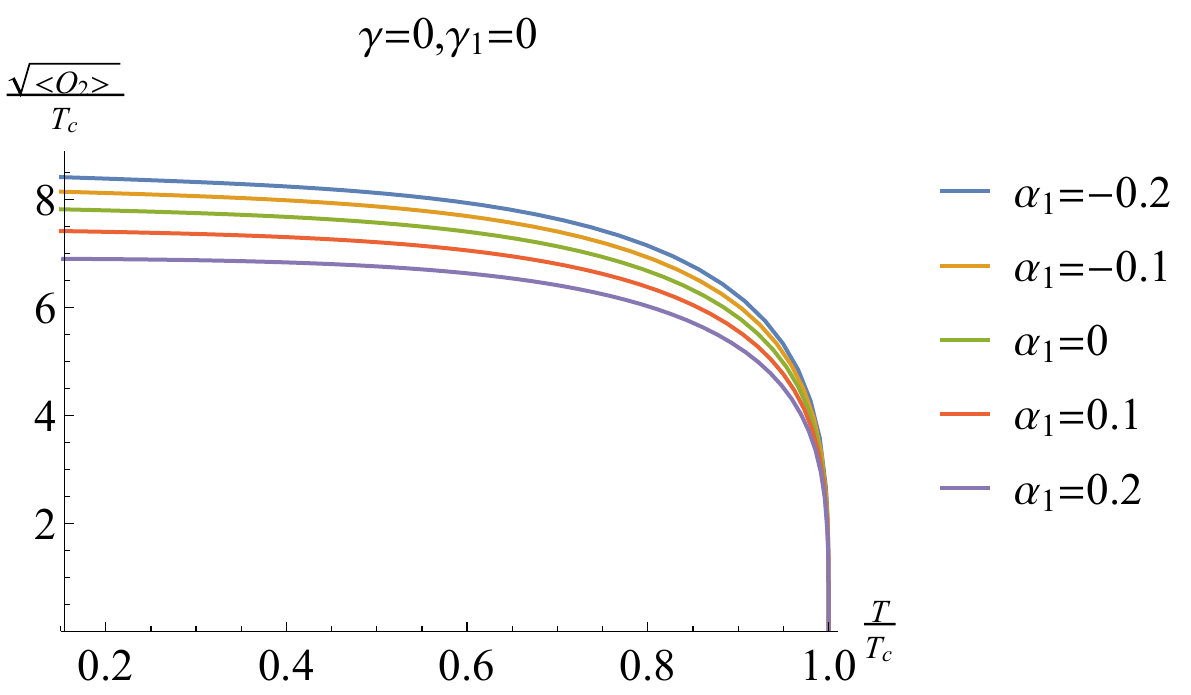}\ \hspace{0.2cm}
\caption{\label{con1} The condensation $\sqrt{\langle O_2\rangle}/T_c$ as a function of the temperature $T/T_c$ for different $\alpha_1$.
Here the mass of scalar field is $m^2=-2$.}}
\end{figure}
FIG.\ref{con1} shows the condensation $\sqrt{\langle O_2\rangle}/T_c$ as a function of the temperature $T/T_c$ for different $\alpha_1$.
We observe that as $\alpha_1$ decreases, the condensation value becomes much larger.
It indicates that a larger superconducting energy gap $\omega_g/T_c$ emerges,
which shall be explicitly addressed in the following study of optical conductivity.
Furthermore, we find that with the increase of the $\alpha_1$, the critical temperature $T_c$ goes up (see TABLE \ref{table-v1}).
The tendency agrees with that shown in the left plot in FIG.\ref{fig-Tcvsalpha1}.
It indicates that the positive $\alpha_1$ term drives the symmetry breaking and makes the condensation easy.
While for the negative $\alpha_1$, the condensation becomes hard and when $\alpha_1$ is less than certain value,
the condensation is spoiled. The result is consistent with the instability analysis in Section \ref{sec-setup}.
\begin{widetext}
\begin{table}[ht]
\begin{center}
\begin{tabular}{|c|c|c|c|c|c|}
         \hline
~$\hat{T_{c}}$~&~$\alpha_1=-0.2$&~$\alpha_1=-0.1$~&~$\alpha_1=0$~&~$\alpha_1=0.1$&~$\alpha_1=0.2$
          \\
        \hline
~$\gamma=1/12$~ & ~$0.0666$~& ~$0.0694$~&~$0.0732$~&~$0.0782$&~$0.0855$
          \\
        \hline
~$\gamma=0$~ &~$0.0545$~& ~$0.0563$~&~$0.0587$~&~$0.0621$&~$0.0670$
          \\
        \hline
~$\gamma=-1/12$~ &~$0.0491$~&~$0.0506$~&~$0.0525$~&~$0.0552$&~$0.0592$
          \\
        \hline
\end{tabular}
\caption{\label{table-v1}The critical temperature $\hat{T}_{c}$ with different $\alpha_{1}$ and $\gamma$.
Here $\gamma_1=0$.}
\end{center}
\end{table}
\end{widetext}
\begin{widetext}
\begin{table}[ht]
\begin{center}
\begin{tabular}{|c|c|c|c|c|}
         \hline
~$\hat{T_{c}}$~ &~$\alpha_1=-0.2$&~$\alpha_1=-0.1$~&~$\alpha_1=0.1$&~$\alpha_1=0.2$
          \\
        \hline
~$\gamma_{1}=0.02$~ & ~$0.0825$& ~$0.0874$~&~$0.1019$& ~$0.1137$
          \\
        \hline
~$\gamma_{1}=0$~ & ~$0.0545$& ~$0.0563$~&~$0.0621$& ~$0.0670$
          \\
        \hline
~$\gamma_{1}=-0.02$~ & ~$0.0496$&~$0.0511$~&~$0.0556$& ~$0.0595$
          \\
        \hline
\end{tabular}
\caption{\label{table2}The critical temperature $\hat{T}_{c}$ with different $\alpha_{1}$ and $\gamma_{1}$.
Here $\gamma=0$.}
\end{center}
\end{table}
\end{widetext}

\begin{figure}
\center{
\includegraphics[scale=0.65]{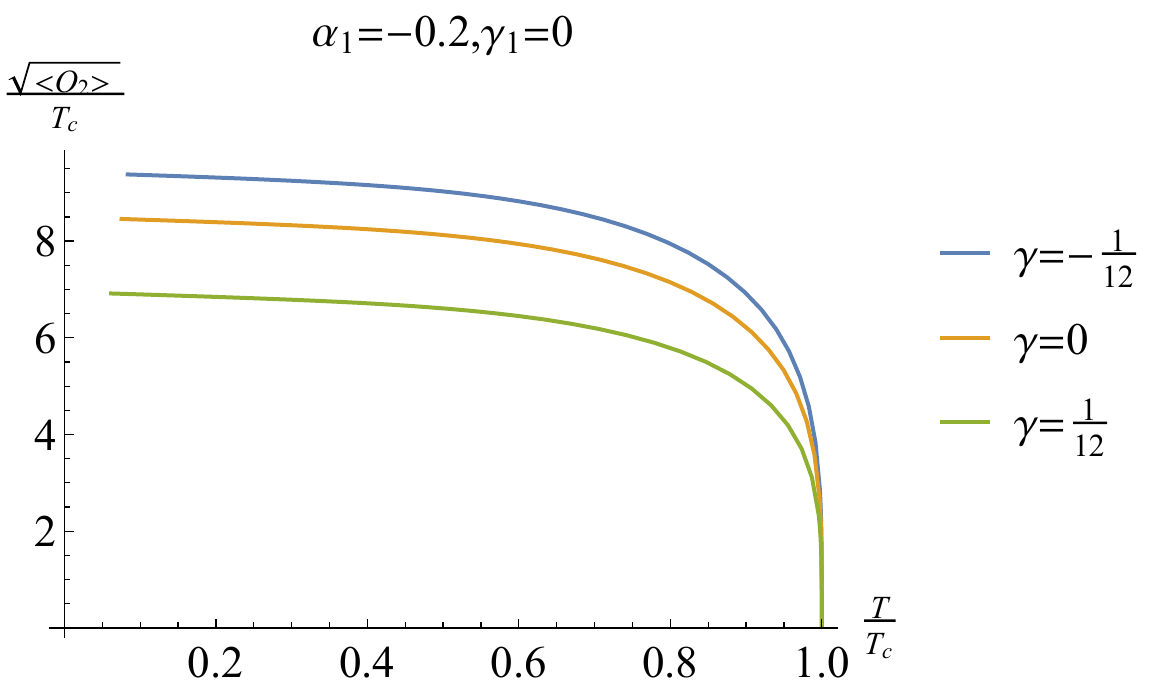}\ \hspace{0.2cm}
\includegraphics[scale=0.65]{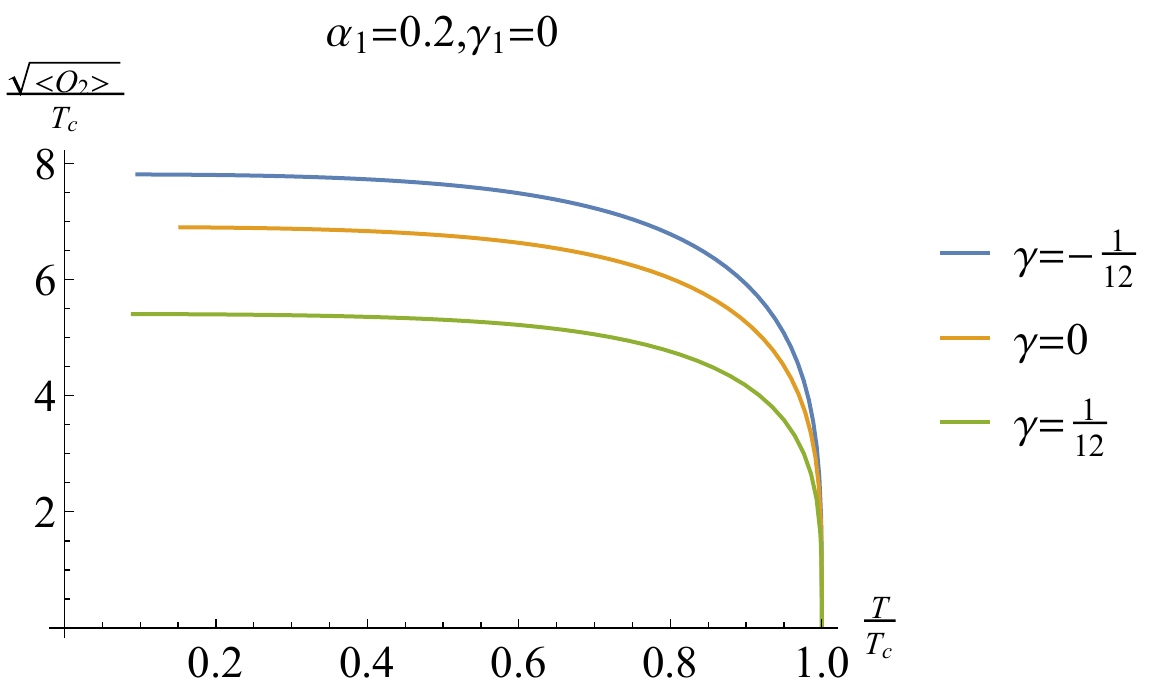}\ \hspace{0.2cm}\\
\includegraphics[scale=0.65]{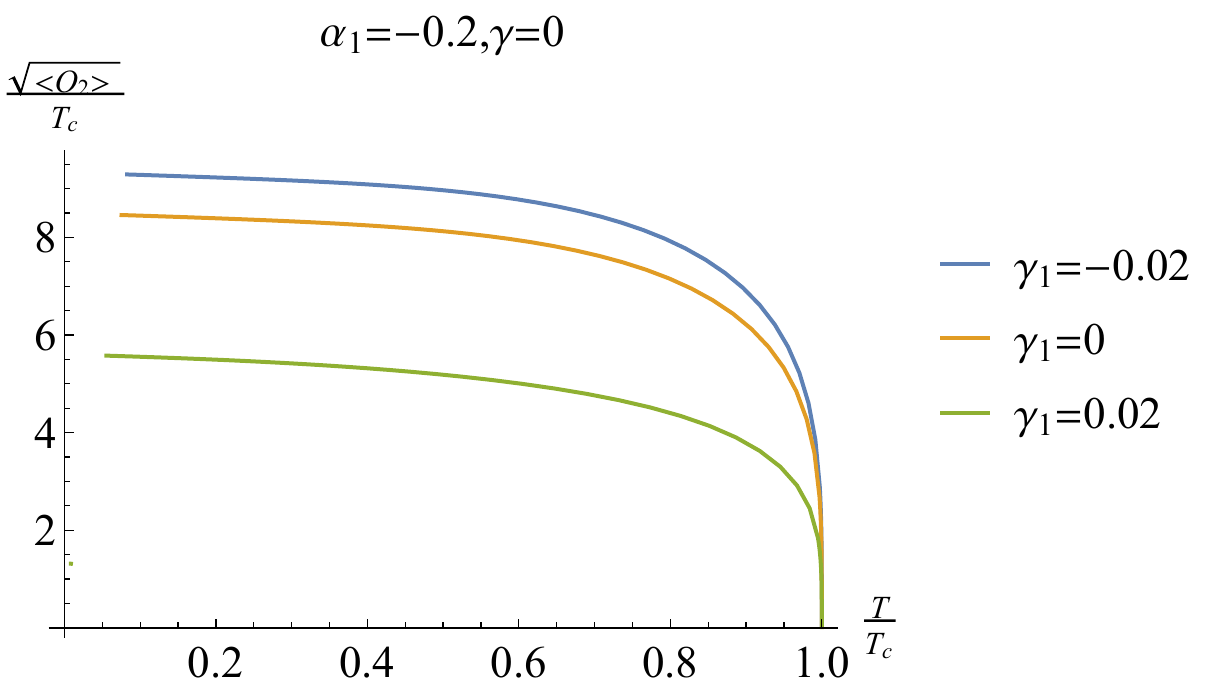}\ \hspace{0.2cm}
\includegraphics[scale=0.65]{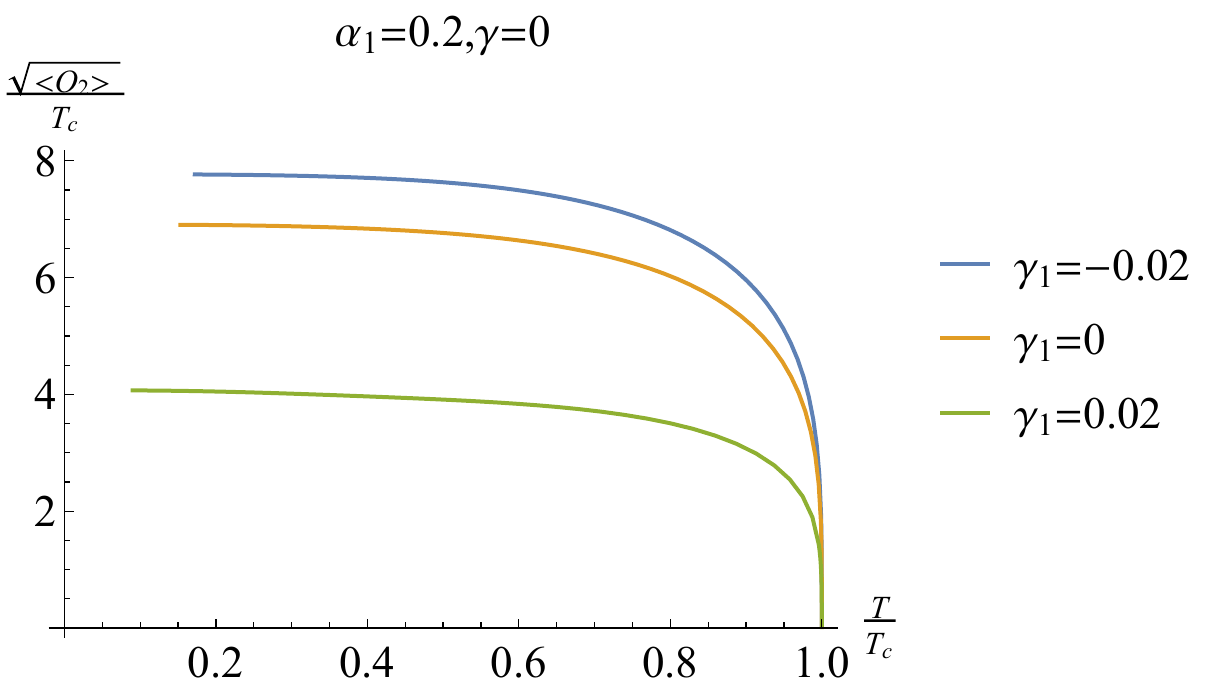}\ \hspace{0.2cm}
\\
\caption{\label{con2} The condensation $\sqrt{\langle O_2\rangle}/T_c$ as a function of the temperature $T/T_c$ for different $\alpha _{1}$ and $\gamma$ or $\gamma_1$.
Here the mass of scalar field is $m^2=-2$.}}
\end{figure}

In previous works \cite{Wu:2010vr,Wu:2017xki}, we have studied the holographic superconductor
from the coupling between Weyl tensor and gauge field.
We find that the condensation value at low temperature runs
and the critical temperature $\hat{T}_c$ changes as the coupling parameters $\gamma$ or $\gamma_1$.
The tendency is consistent with that from the instability analysis in Section \ref{sec-setup}.
Here we further explore the joint effect from $\alpha_1$ term and $\gamma$ or $\gamma_1$ term.
The results are exhibited in TABLE \ref{table-v1}, TABLE \ref{table2} and FIG.\ref{con2}.
We find that the couplings between $\alpha_1$ and $\gamma$ (or $\gamma_1$) have an enhancement or a competitive effect
on the formation of superconducting phase depending on the sign of the coupling parameters.
We summarize the properties as what follows.
\begin{itemize}
  \item
  From TABLE \ref{table-v1} and the panels above in FIG.\ref{con2}, we see that
  both $\alpha_1$ and $\gamma$ are positive, the critical temperature of the superconducting phase transition is enhanced.
  If both $\alpha_1$ and $\gamma$ are negative, the critical temperature is reduced.
  It means that the same sign of $\alpha_1$ and $\gamma$ enhances or reduces the formation of the superconducting phase transition.
  However, if the signs of $\alpha_1$ and $\gamma$ are opposite, there is a competitive effect on the formation of superconducting phase transition.
  \item The same analysis from TABLE \ref{table2} and the panels below in FIG.\ref{con2} gives that when the signs of $\alpha_1$ and $\gamma_1$ are opposite,
  the formation of the superconducting phase transition is enhanced or reduced, while for the same signs of $\alpha_1$ and $\gamma_1$,
  there is a competitive effect on the formation of superconducting phase transition.
\end{itemize}

\section{Conductivity}\label{sec5}

In this section, we study the optical conductivity of our holographic system with HD terms.
For this, we turn on the perturbation of the gauge field along
$x$ direction as $\delta A_x(t,u)=e^{-i\omega t}A_x(u)$.
And then we can derive the following perturbation equation
\begin{align}\label{ax}
(-1+u^{3})(-1-4u^{3}\gamma +48u^{6}\gamma_{1})A_{x}^{''}+(-1+u^{3})(3u^{2}(-1+4\gamma-8u^{3}\gamma -96u^{3}\gamma_{1}+144u^{6}\gamma_{1}))A_{x}^{'}
\nonumber
\\
+((-1-4u^{3}\gamma+48u^{6}\gamma_{1})\omega^{2}-2q^{2}(-2+u^{3})\psi^{2})A_{x}=0.
\end{align}
We can numerically solve the above equation with the ingoing boundary condition.
Once the solution is at hand, we can read off the conductivity in terms of
\fa
\sigma(\omega)=\frac{\partial_u A_x}{i\omega A_x}\Big |_{u=0}\,.
\ffa

\begin{figure}
\center{
\includegraphics[scale=0.65]{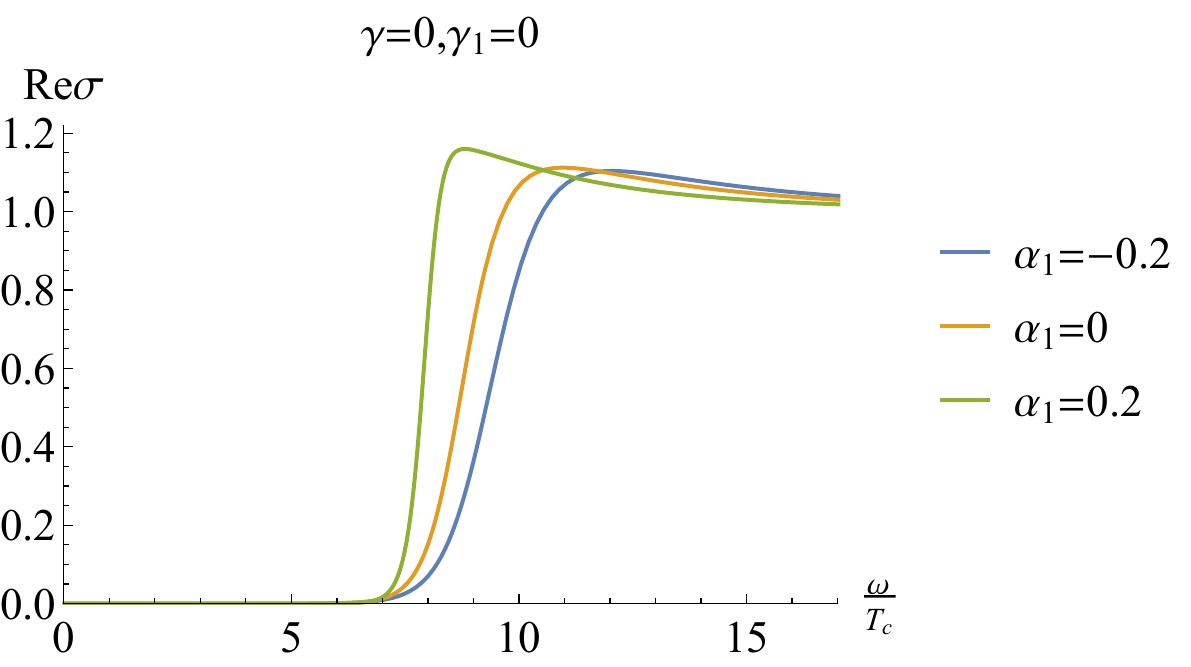}\ \hspace{0.2cm}
\includegraphics[scale=0.65]{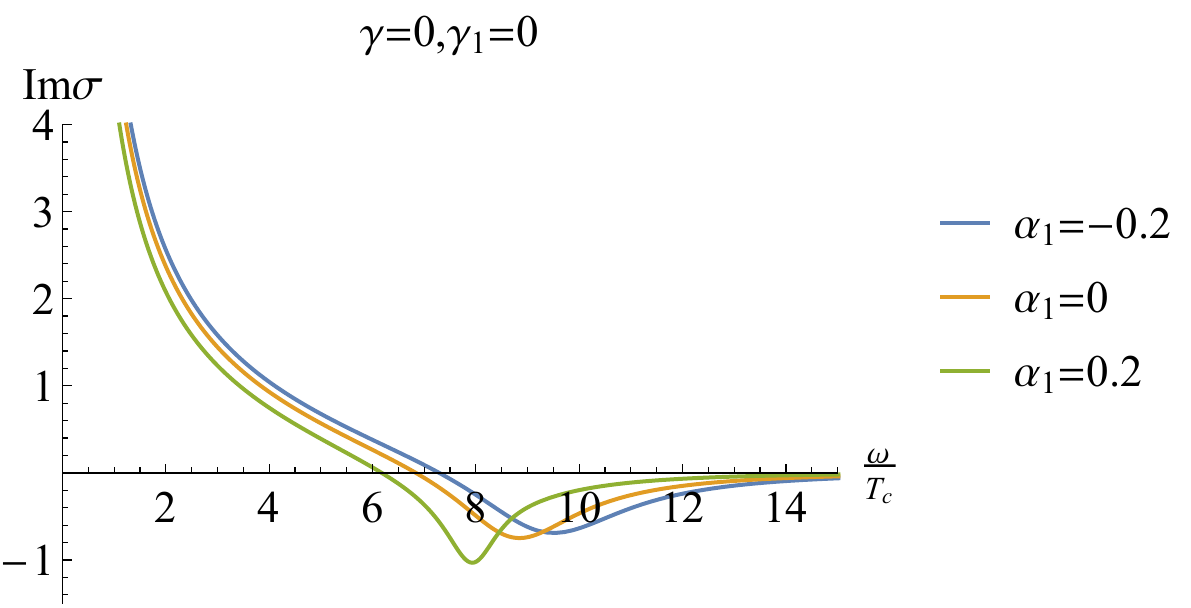}\ \hspace{0.2cm}\\
\caption{\label{con4} Real and imaginary parts of conductivity as a function of the frequency for different $\alpha _{1}$.}}
\end{figure}

\begin{figure}
\center{
\includegraphics[scale=0.65]{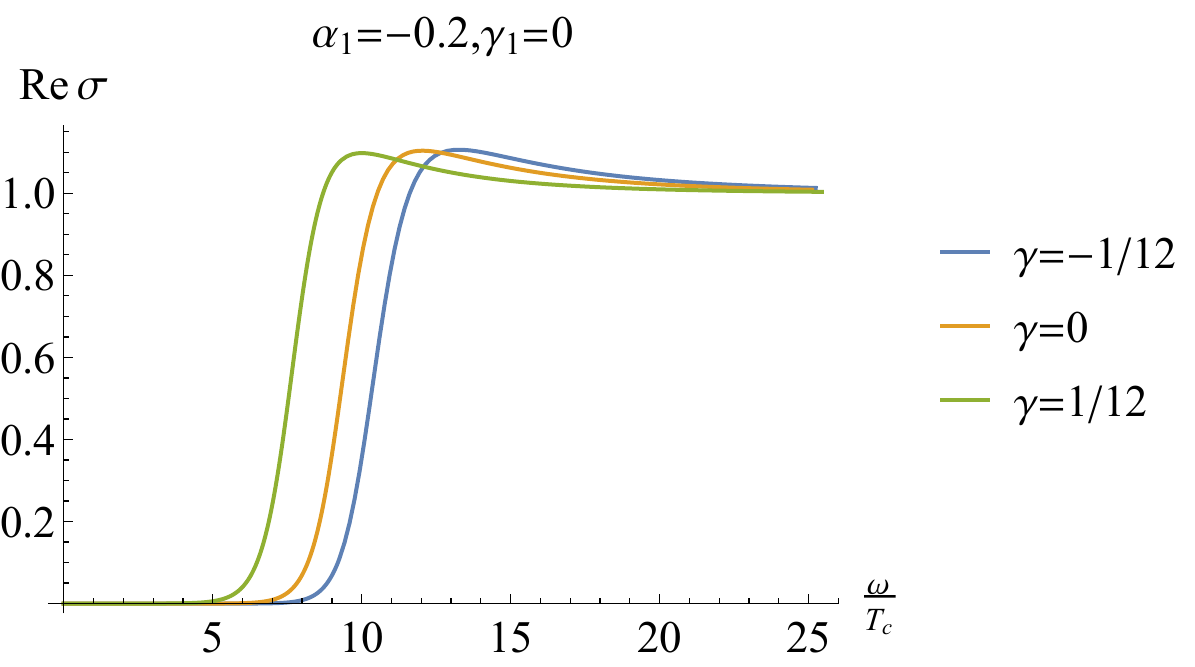}\ \hspace{0.2cm}
\includegraphics[scale=0.65]{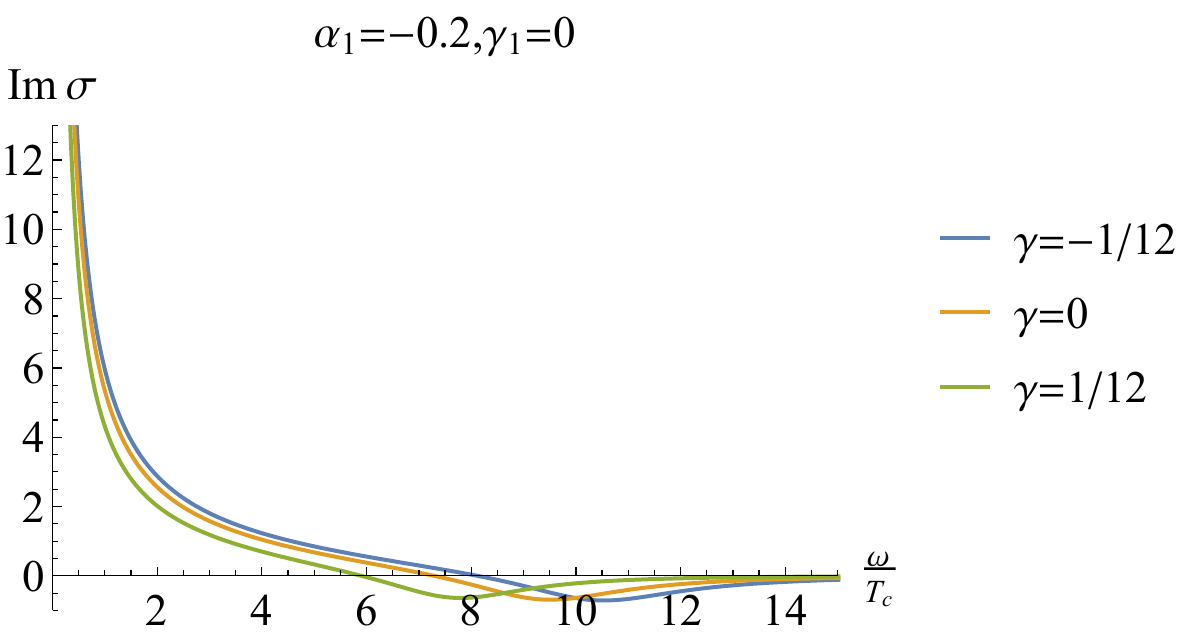}\ \hspace{0.2cm}
\includegraphics[scale=0.65]{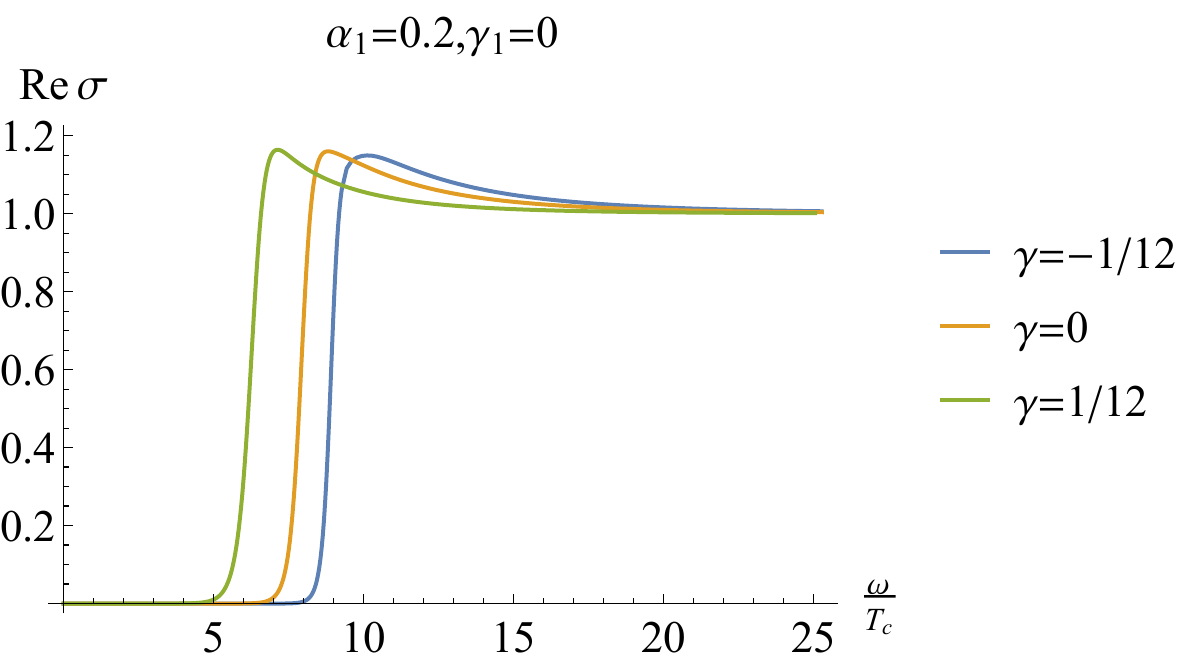}\ \hspace{0.2cm}
\includegraphics[scale=0.65]{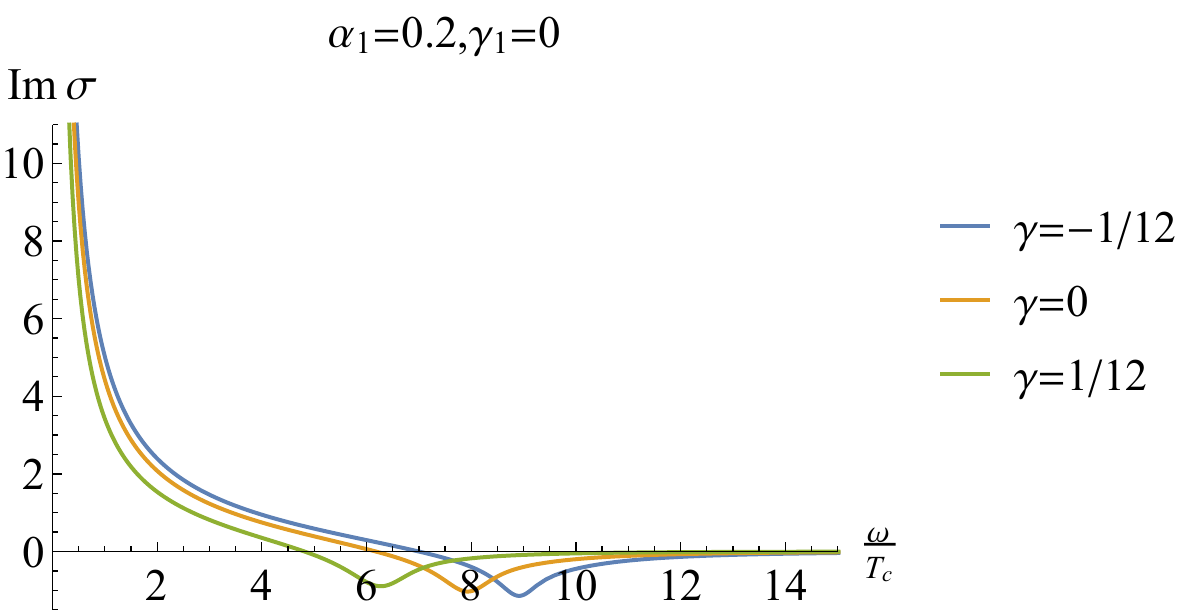}\ \hspace{0.2cm}
\caption{\label{con5} Real and imaginary parts of conductivity as a function of the frequency for different $\gamma$.}}
\end{figure}

\begin{figure}
\center{
\includegraphics[scale=0.65]{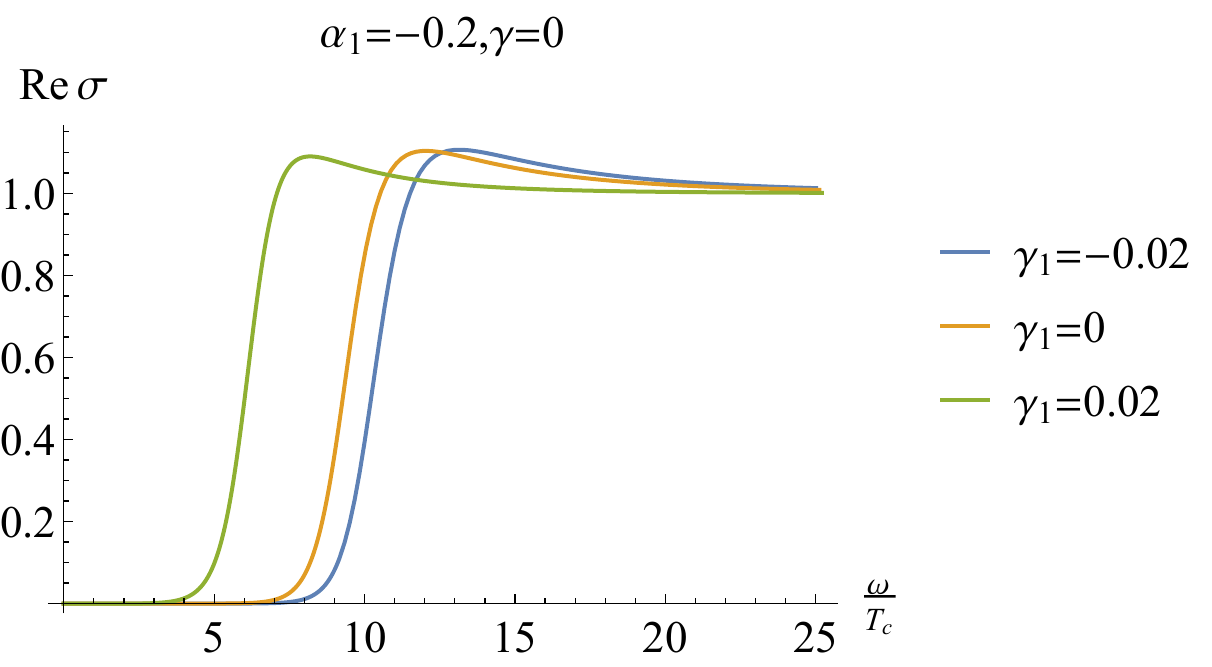}\ \hspace{0.2cm}
\includegraphics[scale=0.65]{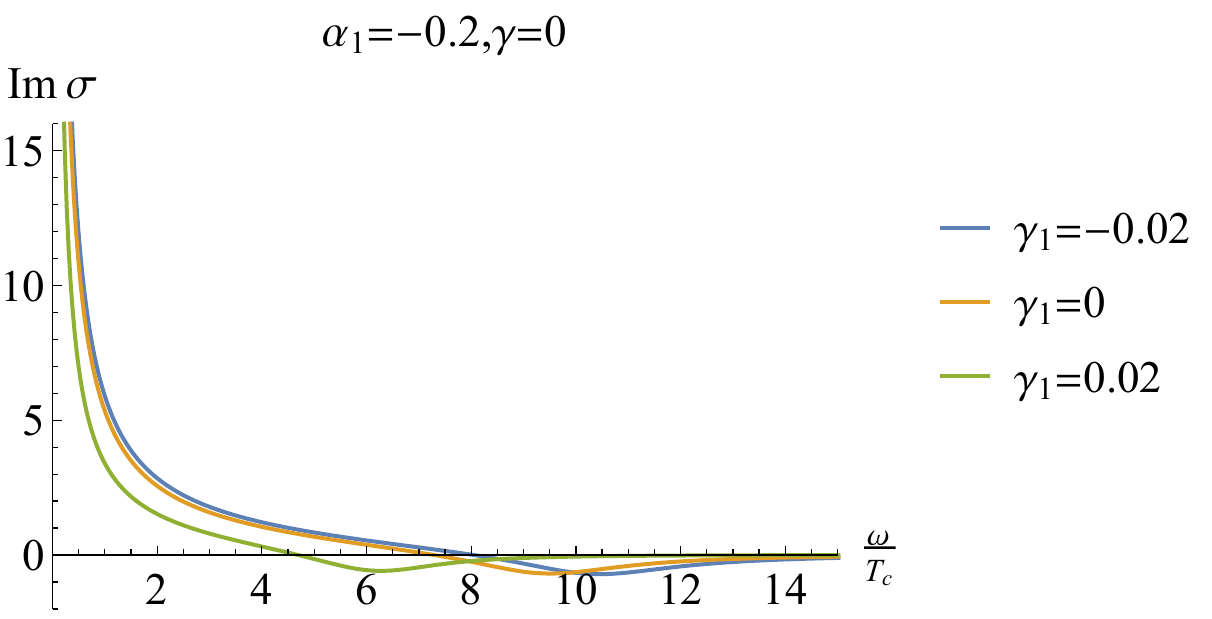}\ \hspace{0.2cm}
\includegraphics[scale=0.65]{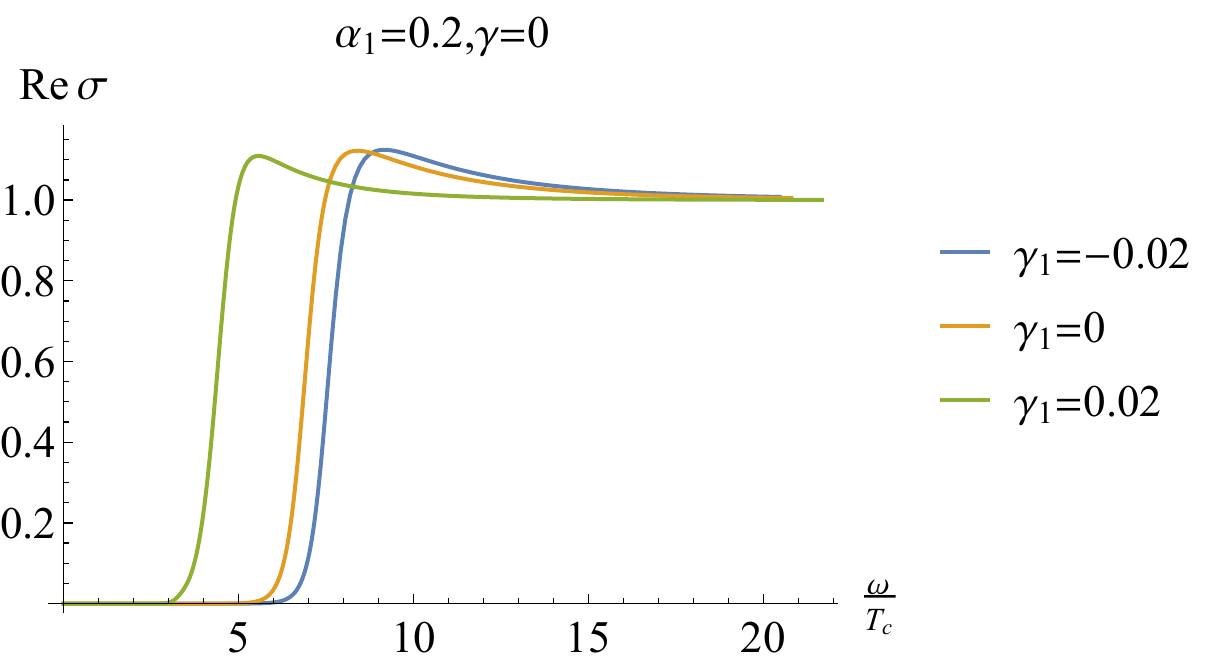}\ \hspace{0.2cm}
\includegraphics[scale=0.65]{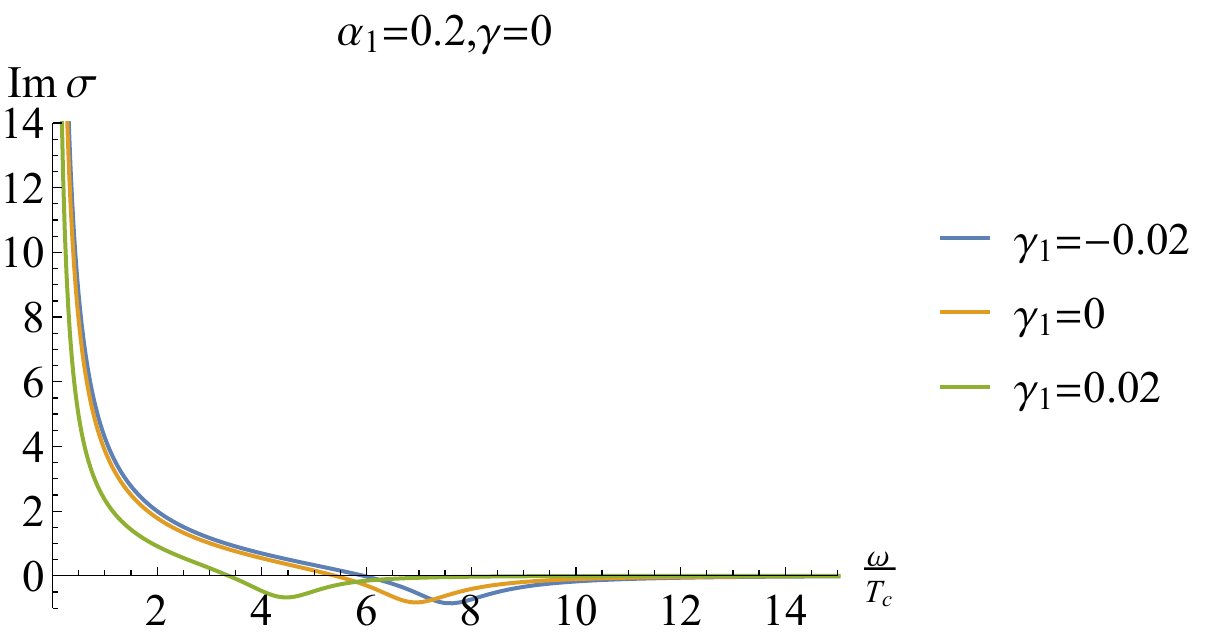}\ \hspace{0.2cm}
\caption{\label{con6} Real and imaginary parts of conductivity as a function of the frequency for different $\gamma_{1}$.}}
\end{figure}

There is no doubt that as the standard version of holographic superconductor model \cite{Hartnoll:2008kx},
the imaginary part of the conductivity (right plots in FIG.\ref{con4}, \ref{con5} and \ref{con6})
has a pole at $\omega=0$, which indicates that there is a delta function at $\omega=0$ in the real part of the conductivity according to the Kramers-Kronig (KK) relation. Such a delta function in conductivity means the emergence of superconductivity.
The another important property that the superconducting energy gap is also clearly exhibited in the real part of the conductivity
(see the left plots in FIG.\ref{con4}, \ref{con5} and \ref{con6}). Near the gap frequency $\omega_g$,
the real part of conductivity quickly goes up, which approximately corresponds to the minimum value in the imaginary part of the conductivity.

In previous works \cite{Wu:2010vr,Ma:2011zze,Mansoori:2016zbp,Wu:2017xki}, the authors have found that after introducing the coupling between the Weyl tensor and the gauge field, the ratio of the superconducting energy gap frequency over critical temperature $\omega_g/T_c$ ranges from about $5.5$ to $16.2$. Here we would also like to explore the running of the superconducting energy gap after the coupling term $\alpha_1$, which is the coupling between the complex scalar field and the Weyl tensor.
From FIG.\ref{con4}, we clearly see that the superconducting energy gap runs with the parameter $\alpha_1$.
Quantitatively, it ranges from about $7.9$, which is less than the value of the standard version holographic superconductor in \cite{Hartnoll:2008kx}, to $9.5$, which is beyond the value of the standard version holographic superconductor, when $\alpha_1\in[-0.2,0.2]$ (see TABLE \ref{table3}).

Furthermore, we also study the the joint effect on the running of the superconducting energy gap from $\alpha_1$ term and $\gamma$ or $\gamma_1$ term.
The results are exhibited in FIG.~\ref{con5},~\ref{con6} and TABLE \ref{table3}, \ref{table4}.
The enhancement or competitive effect on the running of the superconducting energy gap is similar with that of the formation of the superconducting phase discussed in the previous section.
We present a brief summary as what follows.
\begin{itemize}
  \item
  The same sign of $\alpha_1$ and $\gamma$ enhances the running of the superconducting energy gap.
  However, if the signs of $\alpha_1$ and $\gamma$ are opposite, there is a competitive effect on the running of superconducting energy gap.
  \item
  When the signs of $\alpha_1$ and $\gamma_1$ are opposite,
  the running of the superconducting energy gap is enhanced, while for the same signs of $\alpha_1$ and $\gamma_1$,
  there is a competitive effect on the running of superconducting energy gap.
  \item
  The enhancement effect leads to the result that the running range of the superconducting energy gap becomes larger, from $4.6$ to $10.5$
(see FIG.~\ref{con5},~\ref{con6} and TABLE \ref{table3}, \ref{table4}).
\end{itemize}
The extension of the energy gap in our holographic model maybe provide a novel platform to model the high temperature superconductor
and we pursuit it in future.
\begin{widetext}
\begin{table}[ht]
\begin{center}
\begin{tabular}{|c|c|c|c|c|c|}
         \hline
~$\omega_{g}/T_{c}$~ &~$\alpha_1=-0.2$&~$\alpha_1=-0.1$~&~$\alpha_1=0$~&~$\alpha_1=0.1$&~$\alpha_1=0.2$
          \\
        \hline
~$\gamma=1/12$~& ~$7.7857$ & ~$7.5090$~&~$7.1291$~&~$6.7364$& ~$6.6207$
          \\
        \hline
~$\gamma=0$~ & ~$9.5446$& ~$9.2161$~&~$8.8476$~&~$8.4135$& ~$7.9274$
          \\
        \hline
~$\gamma=-1/12$~& ~$10.6011$ &~$10.2622$~&~$9.9057$~&~$9.3502$& ~$8.9090$
          \\
        \hline
\end{tabular}
\caption{\label{table3}The superconducting energy gap $\omega_{g}/T_{c}$ with different $\alpha_{1}$ and $\gamma$.
 }
\end{center}
\end{table}
\end{widetext}
\begin{widetext}
\begin{table}[ht]
\begin{center}
\begin{tabular}{|c|c|c|c|c|}
         \hline
~$\omega_{g}/T_{c}$~ &~$\alpha_1=-0.2$&~$\alpha_1=-0.1$~&~$\alpha_1=0.1$&~$\alpha_1=0.2$
          \\
        \hline
~$\gamma_{1}=0.02$~ & ~$6.2608$& ~$5.9211$~&~$5.1523$& ~$4.6869$
          \\
        \hline
~$\gamma_{1}=0$~ & ~$9.5446$& ~$9.2161$~&~$8.4135$& ~$7.9274$
          \\
        \hline
~$\gamma_{1}=-0.02$~ & ~$10.5032$&~$10.1741$~&~$9.4234$& ~$8.9416$
          \\
        \hline
\end{tabular}
\caption{\label{table4}The superconducting energy gap $\omega_{g}/T_{c}$ with different $\alpha_{1}$ and $\gamma_{1}$.
 }
\end{center}
\end{table}
\end{widetext}

\section{Conclusion and discussion}\label{sec7}

In this paper, we construct a novel holographic superconductor from HD gravity,
for which we introduce a coupling between the complex scalar field and the Weyl tensor.
The $\alpha_1$ coupling term provides a near horizon effective mass squared but doesn't modify the boundary AdS$_4$ BF bound.
The instability analysis indicates that a quantum phase transition can be triggered by tuning the coupling parameter $\alpha_1$.
In particular, even for the positive mass squared, the superconducting phase transition also happens by tuning the $\alpha_1$.
Therefore, the $\alpha_1$ HD term plays the role of driving the symmetry breaking and results in the superconducting phase transition.
We also explore the instability from the HD terms $\alpha_1$ and $\gamma$, which involves the coupling between the Weyl tensor and gauge field. $\gamma$ coupling term modifies the near horizon AdS$_2$ curvature and thus the near horizon effective mass squared,
which provides a mechanism to result in the superconducting phase transition.

The properties of the condensation and the conductivity in our holographic model are also studied.
Although in the probe limit, we have the same tendency as the instability analysis.
We summarize the main properties of our present model as what follows:
\begin{itemize}
  \item For the positive $\alpha_1$, the condensation becomes easy.
While for the negative $\alpha_1$, the result is just opposite.
  \item A wider extension of the superconducting energy gap, ranging from $4.6$ to $10.5$, is observed.
  We expect that our model provides a novel platform to model and interpret the phenomena in the real materials of
  high temperature superconductor.
\end{itemize}

As a novel mechanism, there are some interesting topics deserving further pursuit in future.
\begin{itemize}
  \item We want to explore whether there is the phenomena of the phase space folding as Ref. \cite{Kim:2009kb} due to the introduction of $\alpha_1$ term.
  \item The Homes' law can be observed in the holographic superconductor model from HD theory \cite{Wu:2017xki},
  which involves the coupling between Weyl tensor and the gauge field.
  It is interesting to explore this issue in our present model.
  \item The transports at full momentum and energy spaces can provide far deeper insights into the holographic system than that at the zero momentum \cite{WitczakKrempa:2013ht,Amado:2009ts}. In future, we shall study this issue in the framework of our present model.
  \item It is also interesting to study the running of the superconducting energy gap in other HD holographic superconductor model, for example \cite{Kuang:2013oqa}.
\end{itemize}

\begin{acknowledgments}
This work is supported by the National Natural Science Foundation of China under
Grant Nos. 11775036, 11847313, 11835009, 11875102, 11975072, and 11690021. J. P. Wu is also supported by Top Talent Support Program from Yangzhou University.
\end{acknowledgments}

\end{document}